\documentclass[pra,superscriptaddress,nofootinbib,aps]{revtex4-1}
 \usepackage[utf8]{inputenc}
 \usepackage[greek,english]{babel}
 \usepackage{bm}
 \usepackage{verbatim}
 \usepackage{amsmath}
 \usepackage{amssymb}
 \usepackage{amsthm}
 \usepackage{latexsym}
 \usepackage{amsfonts}
 \usepackage{epsfig}
 \usepackage{epstopdf}
 \usepackage{wasysym} 
 \usepackage{subfigure}
 \usepackage{color}
 \definecolor{darkblue}{rgb}{0,0,.5}
 \definecolor{BLUE}{rgb}{0,0,1}
 \definecolor{BLACK}{rgb}{0,0,0}
 \usepackage[linktocpage, colorlinks=true, linkcolor=darkblue, citecolor=darkblue]{hyperref}
 \usepackage[all]{hypcap}
 \usepackage[makeroom]{cancel}

\usepackage{easybmat}

\usepackage{dsfont}

\newcommand{\bra}[1]{\langle #1|}
\newcommand{\ket}[1]{|#1\rangle}
\newcommand{\braket}[2]{\langle #1|#2 \rangle}

\newcommand{\hc}{\hat{c}}

\begin{document}

\title{Approximations based on density-matrix embedding theory for density-functional theories}
    \author{Iris Theophilou}
    \email{iris.theophilou@mpsd.mpg.de}
	\affiliation{Max Planck Institute for the Structure and Dynamics of Matter, Center for Free Electron Laser Science, Luruper Chaussee 149, 22761 Hamburg, Germany}
	\author{Teresa E.~Reinhard}
	\email{teresa@dribia.com}
	\affiliation{Max Planck Institute for the Structure and Dynamics of Matter, Center for Free Electron Laser Science, Luruper Chaussee 149, 22761 Hamburg, Germany}
	\affiliation{Dribia Data Research S.L., Carrer Llacuna 162, Planta 3 mòdul 303, 08018 BARCELONA, Spain}
	\author{Angel Rubio}
	\email{angel.rubio@mpsd.mpg.de}
	\affiliation{Max Planck Institute for the Structure and Dynamics of Matter, Center for Free Electron Laser Science, Luruper Chaussee 149, 22761 Hamburg, Germany}
	\affiliation{Center for Computational Quantum Physics (CCQ), Flatiron Institute, 162 Fifth Avenue, New York NY 10010, USA}
	\author{Michael Ruggenthaler}
	 \email{michael.ruggenthaler@mpsd.mpg.de}
	\affiliation{Max Planck Institute for the Structure and Dynamics of Matter, Center for Free Electron Laser Science, Luruper Chaussee 149, 22761 Hamburg, Germany}

\date{\today}%

\begin{abstract}
Recently a novel approach to find approximate exchange-correlation functionals in density-functional theory was presented (U. Mordovina et. al., Journal of Chemical Theory and Computation 15, 5209 (2019)), which relies on approximations to the interacting wave function using density-matrix embedding theory (DMET). This approximate interacting wave function is constructed by using a projection determined by an iterative procedure that makes parts of the reduced density matrix of an auxiliary system the same as the approximate interacting density matrix. If only the diagonal of both systems are connected this leads to an approximation of the interacting-to-non-interacting mapping of the Kohn-Sham approach to density-functional theory. Yet other choices are possible and allow to connect DMET with other density-functional theories such as kinetic-energy density functional theory or reduced density-matrix functional theory. In this work we give a detailed review of the basics of the DMET procedure from a density-functional perspective and show how both approaches can be used to supplement each other. We do so explicitly for the case of a one-dimensional lattice system, as this is the simplest setting where we can apply DMET and the one that was originally presented. Among others we highlight how the mappings of density-functional theories can be used to identify uniquely defined auxiliary systems and auxiliary projections in DMET and how to construct approximations for different density-functional theories using DMET inspired projections. Such alternative approximation strategies become especially important for density-functional theories that are based on non-linearly coupled observables such as kinetic-energy density-functional theory, where the Kohn-Sham fields are no longer simply obtainable by functional differentiation of an energy expression, or for reduced density-matrix functional theories, where a straightforward Kohn-Sham construction is not feasible. 
\end{abstract}

\maketitle


\section{Introduction}\label{section:intro}
Finding the ground state of a multi-electron system is of central importance in several areas of modern physics. Yet the exponential increase of the dimension of the interacting multi-electron wave function prohibits a direct solution of the resulting Schrödinger equation in most cases. A possible way to avoid this problem is to reformulate the multi-electron problem in terms of reduced quantities that can be calculated numerically efficiently. Most prominent is density-functional theory~\cite{parr} and its extensions such as one-body reduced density-matrix functional theory~\cite{Pernal2016}. However, the main challenge for density-functional theories is to find accurate yet efficient approximations to the unknown exchange-correlation functionals. Traditionally these functionals are based on approximate energy expressions of simple reference systems such as the homogeneous electron gas~\cite{parr}. It is then necessary to perform a functional derivative with respect to the reduced quantity to obtain the exchange-correlation potentials of the Kohn–Sham approach to density-functional theories.  However, besides fundamental issues with the differentiability of the involved functionals~\cite{Lammert2007,kvaal2014}, it is particularly challenging to construct approximations that also hold for situation with strong static correlations with such energy-based approximation schemes. Therefore alternative approximation strategies are highly desirable. Recently, such an alternative approach was presented in Ref.~\cite{SDE}, where instead of energy expressions directly an approximation to the interacting wave function based on an auxiliary non-interacting wavefunction is employed.  This is done by using ideas from density-matrix embedding theory (DMET)~\cite{DMET}, where an interacting electronic problem is divided into subsystems (referred to as impurity and environment) that are treated on different levels of accuracy. The main connection to density-functional theories and the crucial ingredient of DMET is an approximate projection derived from an auxiliary non-interacting system. This approximate projection is determined by an iterative procedure that makes parts of the reduced density matrix of the auxiliary system the same as the approximate interacting density matrix. If only the diagonal of both systems are connected this leads to an approximation of the interacting-to-non-interacting mapping of standard density-functional theory.

The DMET methodology was first presented and benchmarked for one-dimensional and two-dimensional Hubbard lattices~\cite{DMET} and since then numerous studies and extensions of DMET have been presented on Hubbard lattices~\cite{DET, DMET_honeycomb_Hubbard, DMET_science}. Apart from quantum lattice models it has been also applied to ab-initio Hamiltonians to treat certain molecular~\cite{DMET_molecules1,Inter_bath_molecules} and periodic systems~\cite{DMET_DET_solids,DMET_solids}. Furthermore, different extensions of DMET have been developed to apply the method to time-dependent systems~\cite{j_kretchmer} and excited states~\cite{j_tran}, and to coupled electron-phonon models~\cite{b_sandhoefer,t_reinhard}. Also, finite-temperature systems have already been treated with DMET~\cite{c_sun}. 
Numerical shortcomings of the DMET method can be improved by semi-definite programming~\cite{semi_definite}, projected DMET ~\cite{p_dmet} and multiconfigurational DMET~\cite{multiconfigurational_DMET}.

In this work we want to elucidate the connection between the two mentioned approaches to the multi-electron problem, namely density-functional theories and DMET, and highlight how they can be used to supplement each other. To do so we re-examine the foundations of DMET and provide a comprehensive discussion of the basic ingredients. Since in DMET not only the $M$-particle space is relevant (in contrast to most density-functional theories) we discuss in detail how the different spaces, projectors, Hamiltonians and projected Hamiltonians are connected. We will focus on the simplest setting of DMET, i.e., finite one-dimensional lattices. This together with a focus on the simplest iteration procedures (many different have been proposed in the literature) allows us to highlight several subtle issues. Firstly, by carefully constructing different representations of the electronic Fock space, we show how a Hamiltonian given in terms of global fermionic creation and annihilation operators differs to a representation in terms of local fermionic creation and annihilation operators (Sec.~\ref{sec:single_part_Fock_space}). This is connected to the fact that in an only locally anti-symmetrized basis (as is the case for impurity and environment wave functions) the expansions coefficients need to carry the anti-symmetrization. Furthermore we elucidate how an effective chemical potential arises when a Hamiltonian is projected onto a smaller Fock space, and point out discrepancies with respect to previous works in the projected interaction terms (Sec.~\ref{sec:chemical_potential}). After discussing in detail the different projections employed in DMET, we highlight the appearance of the problem of non-interacting $v$ representability of reduced density-matrix functional theory in the DMET procedure. As a result we find infinitely many non-interacting Hamiltonians with a non-local potential that can be used for the auxiliary projection of the DMET procedure (Sec.~\ref{subsec:arbitraryprojection}). This implies a certain arbitrariness in the iteration procedure and the corresponding iterated approximated projections. Furthermore, we show that making these projections exact by increasing the impurity size to half the full system size (the projector becomes the identity operator on the full Fock space) requires a non-trivial adaption of the standard DMET procedure (Sec.~\ref{sec:ExactEmbeddingExtension}). We then highlight how the arbitrariness of the iteration steps can lead to different fixed points of the DMET procedure without further refinements (Sec.~\ref{subsec:wrongprojector}). This problem can persist also when the full embedded (projected) 1RDMs are made to agree (Sec.~\ref{subsec:environment1RDM}). Since we use a general non-local effective potential we can find a similar problem also for a global (many impurities) iteration (Sec.~\ref{subsec:global}). We here then make a connection with density-functional theories, which provide us with mapping and representability theorems to potentially avoid spurious non-uniqueness and non-representability issues. These theorems suggest to express the exact projection in terms of the auxiliary and an exchange-correlation projection (Sec.~\ref{sec:towardsuniqueness}). Finally we discuss how DMET allows us to approximate density-functional-type mappings and how we can construct approximations for different density-functional theories (Sec.~\ref{sec:towardsdensityfunctionals}). 
\\
\\
To ease access to readers unfamiliar with DMET we provide an extensive supplement where the many different concepts are explained with simple examples.


\section{Theoretical Setting}
\label{sec:Theoretical Setting}

Let us, for simplicity and definiteness, choose in our considerations a finite, one-dimensional lattice system. Since we will be changing Hilbert spaces a lot in the following, let us introduce all of these spaces and how they are connected. At the same time we will also define the Hamiltonians and discuss their representations in the different spaces. Finally we will briefly discuss projections of Hamiltonians onto subspaces and some properties of the 1RDM.

\subsection{From single-particle space to the fermionic Fock space}
\label{sec:single_part_Fock_space}

Following the usual construction of quantum physics, we will start with the single-particle space of $N$ sites, which we denote by
\begin{align}
    h_1 \cong \mathbb{C}^{N}
\end{align}
with the usual inner product and $\cong$ meaning isometrically isomorphic. A Hamiltonian $\hat{h}^{(1)}$ on this space can be represented in the standard (site) basis $\ket{i}$ as a hermitean $N \times N$ matrix $h^{(1)}(i,j) = \braket{i}{\hat{h}^{(1)} j}$. With the $N$ eigenfunctions of this matrix $\braket{i}{\phi_\mu} = \phi_\mu(i)$ and their eigenenergies $\epsilon_\mu$ the Hamiltonian can be equivalently represented as
\begin{align}
h^{(1)}(i,j) = \sum_{\mu=1}^{N} \epsilon_\mu \phi_{\mu}(i) \phi_{\mu}^{*}(j).
\label{eq:single_part_gen}
\end{align}
While we will give several explicit examples for spinless fermions in the supplement (to keep the dimensions small), in general we will consider spin $1/2$ particles. All the results in the following will not depend on whether we include spin or not. The only difference lies in the dimensionalities of the objects that we consider. Since we will keep the spin dimension (a factor 2) explicit, it is usually easy to infer the spinless dimensions (else we state it explicitly). The single-particle space including spin we denote by
\begin{align} 
    \mathcal{H}_1 = h_1 \otimes \mathbb{C}^2 \cong \mathbb{C}^{2N}. 
\end{align}
Here the standard (site-spin) basis is denoted as $\ket{z} \equiv \ket{i \sigma}$ and a Hamiltonian $\hat{H}^{(1)}$ can be represented as a $2N \times 2N$ hermitean matrix that reads in eigenrepresentation 
\begin{align} \label{eq:NonInteracting Hamiltonian}
    H^{(1)}(z,z') = \sum_{\mu=1}^{2N} \epsilon_\mu \phi_{\mu}(z) \phi_\mu^{*}(z').
\end{align}
So far no statistics of the particles have entered our construction. Now for the $M$-particle space the fermionic nature of our electrons will become important. It is common practice to construct the $M$-particle space in two consecutive steps. First we define the space of distinguishable particles as $\mathcal{H}_M = \mathcal{H}_1 \otimes \dots \otimes \mathcal{H}_1$, which has dimensions $(2N)^M$ and standard basis states of the form $|z_1 ... z_M ) = \ket{z_1} \otimes  \dots \otimes \ket{z_M}$. We want to emphasize that we denote the distinguishable-particle (non-symmetrized) basis with $| \cdot )$ while we later denote the indistinguishable-particle (anti-symmetrized) basis with $\ket{\cdot} $. In this space the non-interacting $M$-particle Hamiltonian is defined as $\hat{H}^{(M)} = \hat{H}^{(1)} \otimes \hat{\mathds{1}}^{(1)} \otimes \dots \otimes \hat{\mathds{1}}^{(1)} + \dots + \hat{\mathds{1}}^{(1)} \otimes \dots \otimes \hat{\mathds{1}}^{(1)} \otimes \hat{H}^{(1)}$, where $\hat{\mathds{1}}^{(1)}$ is the identity of $\mathcal{H}_1$. If we denote $\ket{\phi_{\mu_{1}}}\otimes \dots \otimes \ket{\phi_{\mu_M}} = |\mu_1 ... \mu_M)$ it can be expressed as $\hat{H}^{(M)} = \sum_{\mu_{1} \dots \mu_M=1}^{2N} (\epsilon_{\mu_{1}} + \dots + \epsilon_{\mu_{M}})|\mu_1...\mu_M)(\mu_1...\mu_M|$, which with the expression in the standard basis $(z_1...z_M|\mu_1...\mu_M) = \phi_{\mu_1}(z_1) ... \phi_{\mu_M}(z_M)$ leads to the eigenrepresentation in the spin-site basis. At this point one could wonder why we did introduce a space of distinguishable particles, when we anyway want to work with electrons? As we will show below, in quantum physics we often work explicitly in $\mathcal{H}_M$ but restrict then the allowed states to the indistinguishable ones. Nevertheless, we can equivalently work with the Hilbert space of indistinguishable fermions, as we will also show below. Both approaches look formally similar but have some important differences, that we need to highlight for completeness and to avoid subtle errors. The first approach is straightforward. We make all anti-symmetric products for the standard basis 
\begin{align}
    \ket{z_1...z_M} = \frac{1}{\sqrt{M!}} \sum_{\mathfrak{p}} \sigma(\mathfrak{p}) \ket{\mathfrak{p}(z_1)} \otimes \dots \otimes \ket{\mathfrak{p}(z_M)},
    \label{eq:antisym_products}
\end{align}
where the sum goes over all permutations $\mathfrak{p}$ of the $M$ indices and $\sigma(\mathfrak{p})$ denotes whether the permutation is even (+) or odd (-). In a similar manner we can do that for any other basis, e.g., the eigenbasis of the non-interacting Hamiltonian $\hat{H}^{(M)}$ is denoted as $\ket{\mu_1...\mu_M}$. The number of such states is ${2N \choose M}$. If we now look for the eigenstate of $\hat{H}^{(M)}$, however, restricted on this fermionic subspace, we will find all Slater determinants of the non-interacting Hamiltonian, i.e.
\begin{align}
\tilde{\Phi}(z_1...z_M) &= (z_1...z_M\ket{k_1...k_M} \\
&= \frac{1}{\sqrt{M!}} \sum_{\mathfrak{p}} \sigma(\mathfrak{p}) \phi_{\mathfrak{p}(\mu_1)}(z_1)  \dots  \phi_{\mathfrak{p}(\mu_M)}(z_M).   \nonumber
\end{align}
Instead of working in the higher-dimensional space $\mathcal{H}_M$ and then restricting the allowed states, it is also possible to work directly in the properly anti-symmetrized (fermionic) $M$-particle Hilbert space
\begin{align}
    \mathcal{H}_M^{F} = \mathcal{H}_1 \wedge \dots \wedge \mathcal{H}_1 \cong \mathbb{C}^{{2N \choose M}}, 
\end{align}
which is just the span of all the anti-symmetrized states. The Hamiltonian in this space can then be represented by
\begin{align}
    \hat{H}^{(M)}_F = \sum_{\mu_1=1}^{2N}\dots\!\!\!\!\!\!\sum_{\mu_M>\mu_{M-1}}^{2N}\!\!\!\!\!\!\left( \epsilon_{\mu_1}+...+\epsilon_{\mu_M} \right) \ket{\mu_1...\mu_M}\bra{\mu_1...\mu_M}.
\end{align}
That is, in accordance to the smaller dimension the sums with respect to eigenstates are nested, i.e., $\mu_1<\dots <\mu_M$. Furthermore, with respect to the anti-symmetrized spin-basis states $\ket{z_1 \dots z_M}$ the Slater determinants are now
\begin{align}
\Phi(z_1...z_M) &= \braket{z_1...z_M}{\mu_1...\mu_M} \\
&= \sum_{\mathfrak{p}} \sigma(\mathfrak{p}) \phi_{\mathfrak{p}(\mu_1)}(z_1)  \dots  \phi_{\mathfrak{p}(\mu_M)}(z_M).   \nonumber
\end{align}
Since in the following we will work (almost) exclusively with the anti-symmetrized spaces, our Slater determinants will not have the factor $1/\sqrt{M!}$.

Let us next go one step further and relax the fixed number of particles restriction. To this end we construct the Fock space
\begin{align}
    \mathcal{F} = \bigoplus_{M=0}^{2N} \mathcal{H}_{M}^{F} \cong \mathbb{C}^{2^{2N}},
\end{align}
where the Fock-space dimension is determined by the binomial equality $\sum_{M=0}^{2N} {2N \choose M} = 2^{2N}$. In an overloading of symbols we also denote $\ket{z_1...z_M} \equiv \ket{\emptyset}_0 \oplus \dots \ket{z_1...z_M}_M \dots \oplus \ket{\emptyset}_{2N}$, where $\ket{\emptyset}$ is the null vector in the respective spaces and accordingly also $\ket{\Phi} \in \mathcal{F}$. The non-interacting Hamiltonian can be defined straightforwardly by $\hat{H}=\bigoplus_{M=0}^{2N} \hat{H}^{(M)}_{F}$. Yet instead of this expression we would like to use creation $\hat{c}^{\dagger}_z$ and annihilation operators $\hat{c}_z$, which obey the usual anti-commutation relations $\{ \hat{c}_{z}, \hat{c}_{z'}^{\dagger}\} = \delta_{z z'}$ such that $\ket{z_1...z_M} = \hat{c}^{\dagger}_{z_M}...\hat{c}^{\dagger}_{z_1} \ket{0}$, where $\ket{0} \in \mathcal{H}_{0}^{F}$ is the vacuum state. With these we can then define the creation and annihilation operators for the single-particle eigenstates 
\begin{align}\label{eq:OrbitalBasisCreation}
    \hat{\phi}^{\dagger}_\mu = \sum_{z=1}^{2N} \phi_\mu(z)\hat{c}^{\dagger}_{z},
\end{align}
and accordingly for $\hat{\phi}_{\mu}$, which allows us to express
\begin{align}
\label{eq:FockNonInteracting}
    \hat{H}_s=\sum_{\mu=1}^{2N} \epsilon_\mu \; \hat{\phi}^{\dagger}_\mu\hat{\phi}_{\mu} = \sum_{z, z' = 1}^{2N} H^{(1)}(z,z') \; \hat{c}_{z}^{\dagger} \hat{c}_{z'}.
\end{align}
Here the subindex $s$ indicates in analogy to Kohn-Sham theory a non-interacting Hamiltonian. We will later see how to introduce interactions, which is the reason why a direct solution for even just the ground state becomes in practice unfeasible and we need to resort to approximations. Further, for later reference we want to introduce a basis for the Fock space $\mathcal{F}$ by re-labeling as follows (see Supp.~\ref{appendix_single_many} for an explicit example):
\begin{eqnarray}
|F_1\rangle&=&\ket{0}\nonumber\\
|F_2\rangle&=&\hat{c}^{\dagger}_{1\uparrow}\ket{0}\nonumber\\
|F_3\rangle&=&\hat{c}^{\dagger}_{1\downarrow}\ket{0}\nonumber\\
..\nonumber\\
|F_{2^{2N}}\rangle&=&\hat{c}^{\dagger}_{1 \uparrow} \dots \hat{c}^{\dagger}_{N \downarrow}\ket{0}
\label{eq:fock_basis}
\end{eqnarray} 

While we did nothing intricate, this basis makes the anti-symmetry of the space implicit due to the fixed ordering of the creation operators. This implies that the Hamiltonian of Eq.~\eqref{eq:FockNonInteracting} expressed in this basis will look quite different and the anti-symmetry of the fermionic wave functions will be carried over to the expansion coefficients (see Supp.~\ref{appendix_single_many}). Similar problems arise with a different construction for the Fock space, which uses local Fock spaces $\mathcal{F}_{i} \cong \mathbb{C}^{4}$, i.e., $\mathcal{F}_i = \mathrm{span}\{\ket{0}_i,\ket{\uparrow}_i, \ket{\downarrow}_i, \ket{\uparrow \downarrow}_i  \}$, such that $\mathcal{F}'= \bigotimes_{i=1}^{N} \mathcal{F}_i \cong \mathcal{F}$. 
This allows to use a local site-spin basis $\ket{\nu_1}\otimes \dots \otimes \ket{\nu_N} \in \mathcal{F}'$.  Yet again, this basis is \textit{not} explicitly anti-symmetrized~\footnote{The connection between $\mathcal{F}$ and $\mathcal{F}'$ also amounts to fixing an ordering of the $i$ for all objects, e.g., $i_1>\dots >i_M$}. This can also be seen by the \textit{local} creation $\hat{a}^{\dagger}_{i \sigma}$ and annihilation $\hat{a}_{i\sigma}$ operators, which locally anti-commute, i.e., $\{\hat{a}_{i\sigma}^{\dagger},\hat{a}_{i\sigma'}\}=\delta_{\sigma,\sigma'}$, yet when extended to all of $\mathcal{F}'$ for $i\neq i'$ actually commute, i.e., $[a_{i\sigma}^{\dagger}, \hat{a}_{i'\sigma'}] = 0$. 
As a result, the Hamiltonian of Eq.~\eqref{eq:FockNonInteracting} does not take the same form in terms of the local creation and annihilation operators except for special Hamiltonians like next-neighbor hopping (Hubbard) Hamiltonians. The connection follows the Jordan-Wigner transformation $\hat{c}_{i \sigma} = \exp(\textrm{i} \pi \sum_{\sigma'}\sum_{k'<i} \hat{a}^{\dagger}_{k'\sigma'} \hat{a}_{k'\sigma'}) \; \hat{a}_{i\sigma}$ and accordingly for the creation operator. Furthermore it implies that for fermionic wave functions the expansion coefficients in this basis need to carry the missing anti-symmetry. Such an issue will appear later in our considerations when we want to express a fermionic wave function as an impurity and environment tensor product. 

\subsection{Hamiltonian restricted on Fock subspace}
\label{sec:chemical_potential}

Let us next consider the form of the Hamiltonian of Eq.~\eqref{eq:FockNonInteracting} restricted to a subspace of $\mathcal{F}$. We will not consider just any subspace but we choose a different single-particle basis with creation operators $\hat{\varphi}_{\tilde{k}}^{\dagger}$ and an $M-2n$ state $\ket{\tilde{K}}$ 
such that we have
\begin{align}\label{eq:FockSubSpace}
    \mathcal{E} = \textrm{span}\{\ket{\tilde{K}}, \hat{\varphi}_{1}^{\dagger} \ket{\tilde{K}}, \dots , \hat{\varphi}_{4n}^{\dagger}\dots \hat{\varphi}_{1}^{\dagger}\ket{\tilde{K}}\} \cong \mathbb{C}^{2^{4n}}. 
\end{align}
Here we have chosen all $\tilde{\mu}\in \{1,...,4n\}$ such that $\hat{\varphi}_{\tilde{\mu}}\ket{\tilde{K}} = 0$, and for the explicit exmaple in the supplement the number of basis functions $4n$ are $2n$ without spin. The subspace $\mathcal{E}$ is then its own Fock space of lower dimension with the new vacuum state $\ket{\tilde{0}} = \ket{\tilde{K}}$. To determine the Hamiltonian on this subspace we can define a projector onto $\mathcal{E}$ which we denote by $P_{\mathcal{E}}$ and then find $\hat{H}'_s = P_{\mathcal{E}} \hat{H}_s P_{\mathcal{E}}$. We can either do so by labeling the states similarly to Eq.~\eqref{eq:fock_basis} by $\{ \ket{\tilde{F}_1}, \dots, \ket{\tilde{F}_{2^{4n}}} \}$ and have a representation in an ordered basis (see Supp.~\ref{app:ExactProjection}) or we use the representation in terms of the anti-symmetrized Fock-state basis $\ket{\tilde{\mu}_1 \dots \tilde{\mu}_l \tilde{K}}$. In the latter case, using that we only have contributions for equal number of particles and at most one $\tilde{\mu} \neq \tilde{\mu}'$, we find with $H'(\tilde{\mu}, \tilde{\mu}') = \sum_{z_1,z_2} H^{(1)}(z_1,z_2) \varphi_{\tilde{\mu}}^{*}(z_1)\varphi_{\tilde{\mu}'}(z_2)$ and $\Delta \epsilon = \braket{\tilde{K}}{\hat{H}_s \tilde{K}}$   

\begin{align}\label{eq:FockSubSpaceHamiltonian}
    \hat{H}'_s = \sum_{\tilde{\mu}, \tilde{\mu}'=1}^{4n}  H'(\tilde{\mu}, \tilde{\mu}')\hat{\varphi}_{\tilde{\mu}}^{\dagger}\hat{\varphi}_{\tilde{\mu}'} + \frac{\Delta \epsilon}{2n} \hat{\tilde{N}} . 
\end{align}
Here $\Delta \epsilon \hat{\tilde{N}}$, with $\hat{\tilde{N}} = \sum_{\tilde{\mu}} \hat{\varphi}_{\tilde{\mu}}^{\dagger}\hat{\varphi}_{\tilde{\mu}}$ the particle number operator in $\mathcal{E}$, acts as a chemical potential and takes into account the energy due to $\ket{\tilde{K}}$. Alternatively, we could have just used the identity operator on $\mathcal{E}$ and just added $\Delta \epsilon \hat{\mathds{1}}^{\mathcal{E}}$. If we go beyond non-interacting Hamiltonians we usually add a two-particle interaction term of the form $\hat{W}= \sum_{z_1,z_2,z_2,z_1} W^{(2)}(z_1,z_2,z_3,z_4) \hat{c}_{z_1}^{\dagger}\hat{c}_{z_2}^{\dagger} \hat{c}_{z_2} \hat{c}_{z_1}$. We first represent the interaction term in creation and annihilation operators that contain the above $\hat{\varphi}^{\dagger}_{1}$ to $\hat{\varphi}_{4n}^{\dagger}$, which leads to
\begin{align}
    &W^{(2)}(\mu,\nu,\xi, o) \\
    &= \sum_{z_1,z_2 =1}^{2N} \varphi_\mu^{*}(z_1) \varphi_\nu^{*}(z_2) W^{(2)}(z_1,z_2,z_2,z_1) \varphi_\xi(z_2) \varphi_o(z_1). \nonumber
\end{align}
Here $\mu,\nu,\xi,o$ go from $1$ to $2N$. The first $4n$ correspond to the ones used in $\mathcal{E}$ and the ones from $(4n+1)$ to $(2n+1+M)$ build up $\ket{\tilde{K}}$. Next we rearrange the resulting $\hat{W}$ that acts on all of $\mathcal{F}$ in sums that go from $1$ to $4n$ and sums that go from $4n+1$ to $2N$. Since we have a fixed $\ket{\tilde{K}}$ in all our states, the terms that have one index up to $4n$ and the other three are in $(4n+1)$ to $(2N)$ (and vice versa) are zero. The projection on $\mathcal{E}$ thus becomes
\begin{align}
    \hat{W}' &= \sum_{\tilde{\mu},\tilde{\nu}, \tilde{\xi},\tilde{o}=1}^{4n} W^{(2)}(\tilde{\mu},\tilde{\nu}, \tilde{\xi},\tilde{o}) \hat{\varphi}^{\dagger}_{\tilde{\mu}}\hat{\varphi}^{\dagger}_{\tilde{\nu}} \hat{\varphi}_{\tilde{\xi}} \hat{\varphi}_{\tilde{o}} +\frac{\braket{\tilde{K}}{\hat{W} \tilde{K}}}{2n} \hat{\tilde{N}} \nonumber \\
    &+\sum_{\tilde{\mu},\tilde{\xi}=1}^{4n}\left[ \sum_{\nu=4n+1}^{2n+1+M} \left(W^{(2)}(\tilde{\mu},\nu,\nu, \tilde{\xi}) - W^{(2)}(\tilde{\mu},\nu, \tilde{\xi},\nu)  \right. \right. \nonumber
    \\
    & \; \left. \left. + W^{(2)}(\nu,\tilde{\mu},\tilde{\xi},\nu) - W^{(2)}(\nu,\tilde{\mu},\nu,\tilde{\xi})\right) \right] \hat{\varphi}^{\dagger}_{\tilde{\mu}} \hat{\varphi}_{\tilde{\xi}}. 
    \label{eq:FockSubSpaceInteraction}
\end{align}
Let us note here that the terms of the projected interaction that we find here do not agree with the ones presented in, e.g., Eqs.(16) and (17) of Ref.~\cite{DMET_molecules1}.

\subsection{Properties of the one-body reduced density matrix}
\label{subsec:1RDM}

Let us finally comment also on some general properties of the 1RDM that will become important. For any density matrix (mixed state) $\hat{\rho} = \sum_{l} w_l \ket{\Psi_l}\bra{\Psi_l}$ with $\sum_{l}w_l = 1$ and $\ket{\Psi_l} \in \mathcal{F}$, the 1RDM is given by $\gamma(z_1,z_2) = \textrm{Tr} ( \hat{\rho}\hat{c}^{\dagger}_{z_1} \hat{c}_{z_2} ) = \sum_{\mu=1}^{2N} n_\mu \psi_\mu^{*}(z_1)\psi_\mu(z_2)$, where the latter expression is its diagonal representation in terms of the natural occupation numbers $0 \leq n_\mu \leq 1$ and natural orbitals $\psi_{\mu}(z)$. The diagonal provides the particle number $N = \sum_{z=1}^{2N} \gamma(z,z)$ of the density matrix. Of specific interest are here pure states in the $M$-particle sector of $\mathcal{F}$, where one can distinguish between interacting $M$-particle states $\ket{\Psi}$ with usually $0<n_\mu\leq1$ and non-interacting (Slater determinant) wave functions $\ket{\Phi}$ with $n_1 = \dots =n_M =1$ and the rest zero. This implies that the natural orbitals are equivalent to the orbitals of the Slater determinant, e.g., $\braket{\mu_1 \dots \mu_M}{\hat{c}^{\dagger}_{z_1} \hat{c}_{z_2} \mu_1 \dots \mu_M} = \sum_{i=1}^{M} \phi^{*}_{\mu_i}(z_1) \phi_{\mu_i}(z_2)$. Additionally, it also implies that a 1RDM of an interacting system cannot be reproduced by a single Slater determinant. \footnote{Let us point out that there is a simple way to reproduce \textit{any} 1RDM from the ground state of a non-interacting system: One just needs to make all eigenstates degenerate, i.e., in Eq.~\eqref{eq:FockNonInteracting} we choose all $\epsilon_\mu$ the same, and then we can choose an arbitrary sum of Slater determinants as a representative of the degenerate ground-state manifold. However, this "trick" is not useful for any practical purpose as we will discuss later in Sec.~\ref{subsec:environment1RDM}.}


\section{Exact Embeddings via Projections}

The basic idea of DMET is that we divide the system into a part that we treat in detail -- called the impurity -- and a part that while coupled to the impurity is not treated in detail -- called the environment. This division of the system into impurity and environment and the subsequent reformulation of the problem based on this division is called an embedding. While in practice the impurity is changed consecutively and the calculation is repeated such that we have treated all parts of the system in detail, the basic ingredient is the treatment of a single such impurity. In this section, where we discuss how this can be done exactly, we focus on the specific impurity $A$ which is chosen to consist of the sites $i \in \{1, \dots, n\}$ and the rest we denote by $B$. Thus the corresponding spin-site impurity is $A = \{1, \dots ,2n \}$ and accordingly for $B$ such that $\mathcal{H}_1 \cong A \oplus B$.

\subsection{General embedding projections}

The original (undivided) problem is usually to solve a $M$-particle problem on $\mathcal{H}_{M}^{F}$ with a general (usually interacting) Hamiltonian $\hat{H}_{F}^{M}$. For the DMET embedding procedure it then becomes necessary to lift this problem into Fock space. That is, we consider a hermitean Hamiltonian of the form
\begin{align} \label{eq:ExactInteractingHamiltonian}
    \hat{H} =& \sum_{z_1,z_2}H^{(1)}(z_1,z_2)  \; \hat{c}^{\dagger}_{z_1}
\hat{c}_{z_2}\\
&+ \sum_{z_1,z_2} W^{(2)}(z_1,z_2,z_2,z_1)\; \hat{c}_{z_1}^{\dagger}\hat{c}_{z_2}^{\dagger} \hat{c}_{z_2} \hat{c}_{z_1}. \nonumber
\end{align}
We would then like to solve for the ground-state $\ket{\Psi}$ in the $M$-particle sector. Without further simplifications this amounts to a diagonalization of a ${2N \choose M} \times {2N \choose M}$ dimensional matrix, which already for small systems becomes impossible to perform numerically exactly. We would like to reduce this prohibitively large dimensionality. To do so we assume we would know $\ket{\Psi}$ and in a first step make the problem even more intractable by representing it in some Fock-space basis, e.g. 
\begin{align}
    \ket{\Psi} = \sum_{i=1}^{2^{2N}} \Psi_i \ket{F_i}.
    \label{eq:Psi}
\end{align}
Since each $\ket{F_i} = \ket{F_i^{A}} \otimes \ket{F_i^{B}}$, where $\ket{F_i^{A}} \in \mathcal{F}_A \cong \mathbb{C}^{2^{2n}}$ and $\ket{F_i^{B}} \in \mathcal{F}_B \cong \mathbb{C}^{2^{2(N-n)}}$ belong to the impurity $A$ and the environment $B$, respectively, we can re-express the ground state in a new basis
\begin{align}
\ket{\Psi} = \sum_{i=1}^{2^{2n}}\sum_{j=1}^{2^{2(N-n)}} \Psi_{ij} \ket{F_i^{A}}\otimes \ket{F_j^{B}}.
\label{eq:split2}
\end{align}
Of course, since it is a $M$-particle problem most contributions in the full Fock space are zero (see Supp.~\ref{app:svd_fock} for an explicit example). The expansion coefficients $\Psi_{ij}$ are then called the connection matrix between $\ket{F_i^{A}}$ and $\ket{F_j^{B}}$. Alternatively we could also use, e.g., the local basis $\ket{\nu_1}\otimes \dots \otimes \ket{\nu_N}$ to find such a basis for $A$ and $B$, respectively.

We can then in a next step just keep those contributions that are non-zero, re-order and bring Eq.~\eqref{eq:split2} in a diagonal form (see Supp.~\ref{app:svd_fock} for an explicit example). This procedure can be done efficiently with a singular value decomposition (SVD)~\cite{knabner_lineare_2018} of $\Psi_{ij}$. Assuming without loss of generality $n\leq (N-n)$, this leads to
\begin{align}\label{eq:schmidt}
\Psi_{ij} = \sum_{\alpha=1}^{2^{2n}}\sum_{\beta=1}^{2^{2(N-n)}}U_{i\alpha}\Lambda_{\alpha \beta}V_{\beta j}^{\dagger}.
\end{align}
Here, $U_{i\alpha}$ and $V_{\alpha j}^{\dagger}$ are matrix elements of unitary matrices $U \in \mathbb{C}^{2^{2n}}\times\mathbb{C}^{2^{2n}}$ and $V \in \mathbb{C}^{2^{2(N-n)}}\times\mathbb{C}^{2^{2(N-n)}}$, and $\Lambda_{\alpha \beta}$ is a rectangular diagonal $(2^{2n}\times 2^{2(N-n)})$-dimensional matrix with $2n$ real values $\lambda_{\alpha}$ on its diagonal.
Plugging Eq.~\eqref{eq:schmidt} into Eq.~\eqref{eq:split2} then yields
\begin{align}\label{eq:new_basis_1}
\ket{\Psi}  &= \sum_{i=1}^{2^{2n}} \sum_{j=1}^{2^{2(N-n)}} \sum_{\alpha=1}^{2^{2n}}U_{i\alpha}\lambda_{\alpha}V_{\alpha j}^{\dagger}\ket{F_i^{A}} \otimes \ket{F_j^{B}}, \nonumber\\
&= \sum_{\alpha=1}^{2^{2n}}\lambda_{\alpha} \underbrace{\sum_{i=1}^{2^{2n}}U_{i\alpha}\ket{F_i^{A}}}_{= \ket{A_{\alpha}}} \otimes \underbrace{\sum_{j=1}^{2^{2(N-n)}}V_{\alpha j}^{\dagger}\ket{F_j^{B}}}_{=\ket{B_{\alpha}}},\nonumber\\
&= \sum_{\alpha=1}^{2^{2n}}\lambda_{\alpha}\ket{A_{\alpha}} \otimes \ket{B_{\alpha}}.
\end{align}
We have thus decomposed the ground-state wave function into the sum of tensor products of two different sets of wave functions $\ket{A_{\alpha}}$ and $\ket{B_{\alpha}}$. The states $\ket{A_{\alpha}}$ are defined exclusively on the impurity, while the states $\ket{B_{\alpha}}$ are only defined on the environment (see Supp.~\ref{app:svd} for an explicit example). The new states $\ket{B_{\alpha}}$ (which are now only $2^{2n}$ as opposed to $2^{2(N-n)}$ in \eqref{eq:split2}) are then the only ones still considered of the environment $B$ and constitute what is called a bath for the impurity $A$. This construction of the impurity plus the bath is referred to in the DMET literature as the embedded system.
If we next define a subspace of this embedded system $\textrm{span}\{\ket{A_{\alpha}}\otimes \ket{B_{\beta}}  \;| \alpha, \beta \in \{1, \dots ,2^{2n}\}\} \in \mathcal{F}$, which by construction contains the $M$-particle ground state of interest, and define a corresponding projector
\begin{align}\label{eq:projectorFock}
    \hat{P} = \sum_{\alpha, \beta =1}^{2^{2n}} \ket{A_\alpha}\otimes\ket{B_{\beta}}\bra{A_\alpha}\otimes\bra{B_\beta}
\end{align}
we can define a $2^{4n} \times 2^{4n}$ embedded Hamiltonian by
\begin{align}
    \hat{H}'= \hat{P} \hat{H} \hat{P}.
\end{align}
If we now restrict to only the $M$-particle sector and minimize the energy therein we get back the original wave function by construction. We note, however, that it is not a priori clear how many $M$-particle wave functions are in this subspace and it might be non-trivial to sort these wave functions (see also Supp.~\ref{app:exactFockprojection} for an explicit example). Moreover, since the basis is not properly anti-symmetrized, only properly anti-symmetrized coefficients are allowed in the ensuing minimization. All of this implies that even with the exact states $\ket{A_{\alpha}}$ and $\ket{B_{\alpha}}$ this problem might be as hard to solve in practice as the original one if $n$ is not chosen small enough, i.e., $2^{4n} \ll {2N \choose M}$.

\subsection{Embedding projections from non-interacting systems}
\label{sec:embedding_proj_non-int}

In the case that the Hamiltonian is non-interacting, i.e., takes the form of Eq.~\eqref{eq:FockNonInteracting}, and is non-degenerate we can express the embedded Hamiltonian in a more compact and simple form. This is due to the fact that also the ground state takes the much simpler form 
\begin{align}
    \ket{\Phi} =  \left(\sum_{z=1}^{2N}\underbrace{\phi_{1}(z)}_{=C_{z1}}\hat{c}^{\dagger}_z\right) \dots \left(\sum_{z=1}^{2N}\underbrace{\phi_{M}(z)}_{=C_{zM}}\hat{c}^{\dagger}_z\right)  \ket{0}.
\end{align}
If we use for the $(2N \times M)$-dimensional matrix $C_{z\mu}$ the division
\[
	\begin{BMAT}(r)[4pt,0pt,2cm]{cc}{cc}
	2n & \\ 
	2(N-n) &
	\end{BMAT}
    \begin{BMAT}(r)[-2pt,0pt,2cm]{cc}{cc}
      \left\lbrace \vphantom{\rule{1mm}{14pt}} \right. & \\ 
      \left\lbrace \vphantom{\rule{1mm}{14pt}} \right. &
    \end{BMAT}
    \underbrace{\left(
    \begin{BMAT}(e)[2pt,1cm,2cm]{c.c}{c.c}
       C^{A} & \\ 
       C^{E}  &  
    \end{BMAT}
    \right)}_{M} = C,
\] 
and employ a SVD of the impurity submatrix $C_{z\mu}^{A} = \sum_{x=1}^{2n}\sum_{l=1}^{M} U^{A}_{zx} \Lambda_{xl} V^{A \dagger}_{l\mu}$, where $U_A \in \mathbb{C}^{2n} \times \mathbb{C}^{2n}$  and $V^{A}\in \mathbb{C}^{M} \times \mathbb{C}^{M}$, we find
\[
    \begin{BMAT}(r)[4pt,0pt,2cm]{cc}{cc}
      2n &\\ 
      2(N-n) &
    \end{BMAT}
    \begin{BMAT}(r)[-3pt,0pt,3cm]{cc}{cc}
      \left\lbrace \vphantom{\rule{1mm}{18pt}} \right. & \\ 
      \left\lbrace \vphantom{\rule{1mm}{18pt}} \right. &
    \end{BMAT}
    \left(
        \begin{BMAT}(e)[2pt,1cm,2cm]{c.c}{c.c}
       \begin{array}{l l}
      \overbrace{ U\cdot\lambda}^{2n} & \overbrace{0}^{M-2n}
       \end{array} &\\ 
       C^{E}  &  
    \end{BMAT}
    \right) = C\cdot V = \tilde{C}.
\]

Here, due to $\Lambda$ being a rectangular diagonal $2n\times M$ matrix (assuming that $2n \leq M$) with $2n$ entries $\lambda_{x}$ on the diagonal, we find $U\cdot\lambda \in \mathbb{C}^{2n} \times \mathbb{C}^{2n}$ (see Supp.~\ref{app:svd} for an explicit example). Note that we have overloaded the notation again by choosing the same notation for the matrices in the SVD as before in the Fock-space case. The differences (dimensions) should be obvious from the context. The rotation of orbitals that we performed implies that for $\mu \in \{1,...,2n\}$ the corresponding orbitals $\tilde{C}_{z\mu}$ have non-zero entries on $A$ and $B$, while for $\mu \in \{2n+1,...,M\}$ they only have non-zero entries on $B$. Based on these new orbitals we can introduce new creation and annihilation operators. Instead of defining such Fock-space operators for each $\tilde{C}_{zk}$, which would amount to $M$ creation and annihilation operators, we define $4n+(M-2n)$ by further dividing the first $2n$ orbitals into $2n$ that have non-zero values only on $A$ and $2n$ that have non-zero values only on $B$. With the norm on $A$ defined as $\|\tilde{\varphi}_\mu\|_A = (\sum_{z=1}^{2n}|\tilde{\varphi}_\mu(z)|^2)^{1/2}$ as well as on $B$ via $\|\tilde{\varphi}_\mu\|_B = (\sum_{z=2n+1}^{2N}|\tilde{\varphi}_\mu(z)|^2)^{1/2}$ this leads to
\begin{align}\label{eq:neworbitals}
    \varphi_\mu^{A}(z)&= \frac{1}{\|\tilde{\varphi}_\mu\|_A} \tilde{\varphi}_{\mu}(z) \Theta_{+}(2n-z),
    \\
    \varphi_\mu^{B}(z)&= \frac{1}{\|\tilde{\varphi}_\mu\|_B} \tilde{\varphi}_{\mu}(z) \Theta_{+}(z-2n-1), \nonumber
\end{align}
where $\Theta_{+}(z)$ is the Heaviside step function which is $1$ for $z\geq 0$ and zero else. For $\mu>2n$ we have $\tilde{\varphi}_{\mu}(z) = \varphi_{\mu}^{B}(z)$ by construction. Defining with these states 
\begin{align}\label{eq:NewCreation}
    \hat{\varphi}_{\mu}^{A,\dagger} = \sum_{z=1}^{2N} \varphi_{\mu}^{A}(z)\hat{c}^{\dagger}(z)
    \\
    \hat{\varphi}_{\mu}^{B,\dagger} = \sum_{z=1}^{2N} \varphi_{\mu}^{B}(z)\hat{c}^{\dagger}(z)\nonumber
\end{align}
and accordingly $\hat{\varphi}_{\mu}^{\dagger}$ for $\mu>2n$ we can express the non-interacting ground-state wave function as 
\begin{widetext}
\begin{align}\label{eq:SimpleNonInteractingSVD}
    \ket{\Phi} = \left(\|\tilde{\varphi}_1\|_A \hat{\varphi}_{1}^{A,\dagger} + \|\tilde{\varphi}_1\|_B \hat{\varphi}_{1}^{B,\dagger}   \right) \dots \left(\|\tilde{\varphi}_{2n}\|_A \hat{\varphi}_{2n}^{A,\dagger} + \|\tilde{\varphi}_{2n}\|_B \hat{\varphi}_{2n}^{B,\dagger}   \right) \hat{\varphi}_{2n+1}^{\dagger} \dots \hat{\varphi}_{M}^{\dagger} \ket{0}.
\end{align}
\end{widetext}
This leads to $2^{2n}$ terms which are equivalent to the ones from Eq.~\eqref{eq:new_basis_1} for a non-interacting wave function. So we could now re-arrange the sum, express the different states as $\ket{A_{\alpha}}\otimes \ket{B_{\alpha}}$ and identify the corresponding $\lambda_{\alpha}$, which allows us to define a projection of the form of Eq.~\eqref{eq:projectorFock} (see Supp.~\ref{app:svd} for an explicit example). Instead we use that the above defined creation operators of Eq.~\eqref{eq:NewCreation} span a subspace of the form of Eq.~\eqref{eq:FockSubSpace} and the projection thus leads to a Hamiltonian of the form of Eq.~\eqref{eq:FockSubSpaceHamiltonian}. Since the new non-interacting Hamiltonian can be determined as a $4n \times 4n$ matrix of the form $H'_s = C_{\textrm{CAS}}^{\dagger} H_s C_{\textrm{CAS}}$ with the CAS matrix
\begin{widetext}
\begin{equation} 
C_{\textrm{CAS}}=    \begin{BMAT}(r)[4pt,0pt,2cm]{cc}{cc}
      2n & \\ 
      2(N-n) &
    \end{BMAT}
     \begin{BMAT}(r)[-4pt,0pt,4cm]{cc}{cc}
      \left\lbrace \vphantom{\rule{1mm}{27pt}} \right. & \\ 
      \left\lbrace \vphantom{\rule{1mm}{27pt}} \right. &
    \end{BMAT}
\left(
\begin{BMAT}(r){ccc:ccc}{ccc:ccc}
  1 & 0 &\cdots& \varphi^B_1(1) & \varphi^B_2(1)  & \cdots \varphi^B_{2n}(1)\\
  0 & 1 &\cdots& \varphi^B_1(2) & \varphi^B_2(2)&\cdots \varphi^B_{2n}(2) \\
  \vdots & \vdots &\vdots&\vdots  &\vdots  & \vdots\\
  0 & 0 &\cdots& \cdots  &  & \cdots\\
  0 & 0 &\cdots& \cdots & \cdots & \cdots \\
  0 & 0 &\cdots& \varphi^B_1(2N) & \varphi^B_2(2N)  & \cdots\varphi^B_{2n}(2N)\\
\end{BMAT} 
\right)
\label{eq:CAS_matrix}
\end{equation}
\end{widetext}
we see that $H'_s(z_1,z_2) = H_s(z_1,z_2)$ for $z_1$ and $z_2$ restricted to $z\in A$, i.e., on the impurity the Hamiltonian has the same form (see also Supp.~\ref{embedded_system} for an explicit example). Restricting now to the $2n$ particle subspace in the Fock space $\mathcal{E}$ gives back the ground-state wave function of the original problem, provided we also know the form of $\ket{\tilde{0}} \equiv \ket{\tilde{K}}$ in terms of the original Fock space. If we do not know the form of this new vacuum state in terms of the original basis then we at least still get back the wave function on the impurity $A$ since $\ket{\tilde{K}}$ has zero contribution on $A$. We furthermore see that this procedure, in contrast to the one of Eq.~\eqref{eq:projectorFock}, can only work in general for non-interacting problems. The reason being that an interacting wave function consists of (usually) all possible Slater determinants that we can construct and hence we cannot discard any of the original $2N$ orbitals and corresponding creation operators a priori .

Before we move on, let us highlight that there is a very elegant way to obtain the CAS and the corresponding matrix $C_{\textrm{CAS}}$. If we use the previous SVD for $C_{z\mu}$, the 1RDM of the system can be brought into the form
\begin{align}
    \gamma(z_1,z_2)&=\sum_{k=1}^{M} C_{z_1 k}^* C_{k z_2} = \sum_{\mu=1}^{M} \tilde{C}_{z_1 \mu}^* \tilde{C}_{\mu z_2}\\
    &=\sum_{\mu=1}^{2n} \tilde{\varphi}_\mu^{*}(z_1)\tilde{\varphi}_\mu(z_2) + \sum_{\mu=2n+1}^{M} \tilde{\varphi}_\mu^{*}(z_1)\tilde{\varphi}_\mu(z_2). \nonumber
\end{align}
Using that in the sub-matrix $\gamma_{\textrm{env}}(z_1,z_2)$ of $\gamma(z_1,z_2)$, with $z_1$ and $z_2$ in $B\equiv\{2n+1,\dots,2N\}$, only the $\tilde{\varphi}_\mu(z)$ and $\varphi^{B}_\mu(z)$ from Eq.~\eqref{eq:neworbitals} contribute, we find that
\begin{align}
    \gamma_{\textrm{env}}(z_1,z_2) &= \sum_{\mu=1}^{2n} \|\tilde{\varphi}_\mu\|_B^2 \varphi_\mu^{B,*}(z_1)\varphi_\mu^{B}(z_2) 
    \\
    &+ \sum_{\mu=2n+1}^{M} \varphi_\mu^{B,*}(z_1)\varphi_\mu^{B}(z_2). \nonumber
\end{align}
Thus diagonalizing $\gamma_{\textrm{env}}(z_1,z_2)$ and only keeping those eigenfunctions $\varphi_\mu^{B}(z)$ that have eigenvalues (natural occupation numbers) $0< n_\mu^B = \|\tilde{\varphi}_\mu\|_B^2 < 1$ gives us directly the non-trivial entries of the matrix $C_{\textrm{CAS}}$. See Supp.~\ref{embedded_system} for an example of this construction. For later use we define here also the impurity 1RDM $\gamma_{\rm imp}(z_1,z_2)$, which is the sub-matrix of $\gamma(z_1,z_2)$ with $z_1$ and $z_2$ restricted to $A\equiv\{1,\dots,2n\}$. Furthermore, we define
\begin{align}
   \gamma_{\rm emb}(z_1,z_2) &= \underbrace{\sum_{\mu=1}^{2n}\|\tilde{\varphi}_{\mu}\|_A^2 \varphi_{\mu}^{A,*}(z_1)\varphi^{A}(z_2)}_{=\gamma_{\rm imp}(z_1,z_2)} \\
   &+ \sum_{\mu=1}^{2n}\|\tilde{\varphi}_{\mu}\|_B^2 \varphi_{\mu}^{B,*}(z_1)\varphi^{B}(z_2) \nonumber
\end{align}
the embedded 1RDM, which can also be found by calculating the $2n$-particle ground state of the embedded Hamiltonian $\hat{H}'_s$ and excluding the orbitals of $\ket{\tilde{K}} = \ket{\tilde{0}}$ (also called unentangled occupied/core orbitals).  


\section{Mean-field embeddings, self-consistency and the non-interacting $v$- representability issue}

So far we have only given some basic constituents that are part of the DMET procedure. Let us in the following connect them and discuss in more detail the fundamental algorithm of DMET. While there are many flavours available, we want to stick to the essentials and consider the standard choices where 1RDMs are matched in specific ways. To this end will focus on matching 1RDMs locally (on each impurity).\\

\subsection{Mean-field embedding via impurity one-body reduced density matrix}

As said before, we divide our problem in an impurity $A$ and an environment $B$. To find the exact projector to perform the embedding onto $A$ we would first need to solve the original interacting problem of the form of Eq.~\eqref{eq:ExactInteractingHamiltonian}. This is of course not practical because the DMET procedure was developed to avoid exactly this unfeasible numerical task. Hence in the following we want to reduce the dimension of our problem which is ${2N \choose M}$. The goal is now to find an \textit{approximate} projection $\hat{P}$. If we just use any approximate version of the form of Eq.~\eqref{eq:projectorFock} we work in a sub-space of the full Fock space with the dimension $2^{4n}$.  Already at this point we highlight that the moment we assume the size of $A$ to be half the system, i.e., $2n = N$, nothing is discarded and the projector becomes the identity, i.e., we are back in needing to solve the original problem. To find the approximate ground state (due to the approximate projection) we then need to restrict to those states that provide exactly $M$ particles. To identify these states can be cumbersome (see also Supp.~\ref{app:ExactProjection} for an explicit example) and hence it is desirable to have an ordering by particle number a priori. The non-interacting projections provide such an ordering, since they give rise to a new Fock space $\mathcal{E}$ and purpose-built Slater determinants. Hence, in practice a non-interacting projector is used. But instead of just, e.g., the projection from the Eq.~\eqref{eq:ExactInteractingHamiltonian} with $W^{(2)} \equiv 0$, a self-consistency condition is enforced. Which condition and how it is enforced then connects DMET to different density-functional theories. With a non-interacting projector we therefore have the dimension ${4n \choose 2n}$, where we have assumed above that $2n \leq M$ holds. However, if $2n>M$ the dimension becomes ${4n \choose M}$ (which as one could verify gives back the original problem in the limit that $2n=N$). It is important to note here that an adaptation of how the approximate projection is determined in general would be needed for $2n>M$ (see discussion in Sec.~\ref{sec:ExactEmbeddingExtension}).

A standard self-consistency condition is then
\begin{align}\label{eq:SelfConsistency}
    \gamma_{\textrm{imp}}^{s}(z_1,z_2) = \gamma_{\textrm{imp}}'(z_1,z_2),
\end{align}
where $\gamma_{\textrm{imp}}^{s}(z_1,z_2)$ is the 1RDM on the impurity of the auxiliary non-interacting system that provides the approximate mean-field projector $\hat{P}_s$, and $\gamma_{\textrm{imp}}'(z_1,z_2)$ is the 1RDM on the impurity of the projected interacting problem with Hamiltonian $\hat{P}_s \hat{H} \hat{P}_s$ in the respective $M$-particle sector. We note, however, that unless the impurity is half of the system size $2n=N$ (where one solves practically the original problem) it is not guaranteed that $\gamma_{\textrm{imp}}'(z_1,z_2)$ and thus also the approximate interacting wave function is close to the exact $\gamma_{\textrm{imp}}(z_1,z_2)$ and the exact interacting wave function $\ket{\Psi}$.

\subsection{Non-interacting $v$-representability: ambiguities in the mean-field projection}
\label{subsec:arbitraryprojection}

In order to attain self-consistency we need to define the mean-field Hamiltonian which gives the approximate projection iteratively.
As we will show by the following construction there are ambiguities in this procedure as there are infinitely many non-interacting Hamiltonians that reproduce a given impurity 1RDM (see also Supp.~\ref{app_project} for an explicit example). We then discuss the connection of this result to the problem of the non-interacting $v$-representability of 1RDMs. \par

As an initial guess we can, e.g., solve Eq.~\eqref{eq:ExactInteractingHamiltonian} without interaction (although there are different choices). The resulting $\hat{P}^{(0)}_s$ is then used to solve $\hat{P}_s^{(0)} \hat{H} \hat{P}^{(0)}_s$, from which we can determine $\gamma_{\textrm{imp}}^{(0)}(z_1,z_2)$. In a next step a non-interacting system is constructed such that it reproduces the interacting 1RDM submatrix $\gamma_{\textrm{imp}}^{(0)}(z_1,z_2)$ on the impurity $A$.
First we diagonalize on $A$
\begin{align}
  \gamma_{\textrm{imp}}^{(0)}(z_1,z_2) = \sum_{\mu=1}^{2n} \|\tilde{\varphi}_\mu\|_A^2 \varphi_\mu^{A,*}(z_1)\varphi_\mu^A(z_2),  
\end{align}
where we have denoted the corresponding natural occupation numbers and natural orbitals in accordance to Eq.~\eqref{eq:neworbitals}. Now we only need to reverse the steps that led to Eq.~\eqref{eq:neworbitals}. Firstly we choose $2n$ arbitrary states $\varphi_\mu^B(z)$ that are orthonormalized on $B$. Since $B$ has a size of $(2N-2n)$ we have as many choices. With $\|\tilde{\varphi}_\mu\|_B^2 = 1-\|\tilde{\varphi}_\mu\|_A^2$ we then define for $\mu \in \{1,\dots,2n\}$ states
\begin{align}
    \tilde{\varphi}_{\mu}(z) = \|\tilde{\varphi}_\mu\|_A \varphi_\mu^A(z) + (1-\|\tilde{\varphi}_\mu\|_A^2)^{1/2} \varphi_\mu^B(z)
\end{align}
(where which state on $A$ goes together with which state on $B$ is again completely arbitrary).
Since we have assumed $2n<M$ we have to choose $(M-2n)$ further arbitrary orthonormal orbitals $\varphi_\mu^{B}(z)$ (of course orthogonal to the previous $2n$) and define for $\mu \in \{2n+1, \dots , M \}$ states
\begin{align}
    \tilde{\varphi}_{\mu}(z) \equiv \varphi_\mu^B(z).
\end{align}
We have thus constructed $M$ orthonormal single-particle states $\tilde{\varphi}_{\mu}(z)$ with $\mu \in \{1,\dots ,M\}$ on $\mathcal{H}_1$. Since $\mathcal{H}_1$ has a dimension of $2N$, we are left with $(2N-M)$ further orthonormal states that we again order arbitrarily and denote by $\tilde{\varphi}_{\mu}(z)$ for $\mu \in \{M+1, \dots, 2N\}$. As a final step we choose arbitrary energies $\tilde{\epsilon}_\mu \in \mathbb{R}$ such that
\begin{align}
    \tilde{\epsilon}_1 \leq \dots \leq \tilde{\epsilon}_M < \tilde{\epsilon}_{M+1} \leq \dots \leq \tilde{\epsilon}_{2N}.
\end{align}
With these ingredients we find a single-particle Hamiltonian 
\begin{align}\label{eq:AuxiliaryHamiltonian}
    \tilde{H}^{(1)}(z,z') = \sum_{\mu=1}^{2N} \tilde{\epsilon}_\mu \tilde{\varphi}_{\mu}^{*}(z)\tilde{\varphi}_{\mu}(z)
\end{align}
and a corresponding Fock-space Hamiltonian with $\hat{\tilde{\varphi}}_\mu^{\dagger} = \sum_{z=1}^{2N} \tilde{\varphi}_\mu(z)\hat{c}^{\dagger}_{z}$ that has as its $M$-particle ground state $\ket{\tilde{\Phi}} = \hat{\tilde{\varphi}}_M^{\dagger} \dots \hat{\tilde{\varphi}}_1^{\dagger} \ket{0}$. And by construction $\gamma_s(z_1,z_2) = \braket{\tilde{\Phi}}{\hat{c}_{z_1}^{\dagger} \hat{c}_{z_2}\tilde{\Phi}} \equiv \gamma_{\textrm{imp}}^{(0)}(z_1,z_2)$ if restricted to $z_1$ and $z_2 \in A$.

Let us note that we have just shown that there are infinitely many $\tilde{H}^{(1)}(z,z')$ that reproduce a given impurity 1RDM. Except of $\varphi_{\mu}^{A}(z)$ every other part of our construction is completely arbitrary. Yet different choices generate different projections $\hat{P}_s^{(1)}$ and corresponding subspaces $\mathcal{E}^{(1)}$. And if we now proceed with our iteration, each of this projector will lead to a different $\hat{P}_s^{(1)} \hat{H} \hat{P}_s^{(1)}$ and consequently different $\ket{\Psi^{(1)}}$ as well as $\gamma_{\textrm{imp}}^{(1)}(z_1,z_2)$. This is one reason why in practice the iteration might not converge. Such an ambiguity with respect to the non-interacting Hamiltonians is well known in reduced density-matrix functional theories~\cite{theophilou2015orbitals,giesbertz2019one}. It is called the non-interacting $v$-representability problem. It states that a non-interacting 1RDM can be generated by the ground state of many different non-interacting Hamiltonians that differ with regard to their non-local potentials $v$. It stems from the fact that for a non-degenerate non-interacting 1RDM only the first $M$ orbitals are occupied. If we, however, consider a single-particle space of dimension $2N>M$, the rest of the orbitals are not determined and we can thus have many Hamiltonians (see Eq.~\ref{eq:NonInteracting Hamiltonian}) that have the same non-interacting wave function as ground state. This, together with the fact that a non-degenerate non-interacting Hamiltonian cannot reproduce the 1RDM of an interacting system (see Sec.~\ref{subsec:1RDM}), prohibits usually the use of an auxiliary non-interacting system in 1RDM functional theory~\cite{theophilou2015orbitals,giesbertz2019one}. Instead one has to enforce representability conditions of the 1RDMs, which except for ensembles increase exponentially with the dimension of the single-particle space and the number of particles~\cite{Klyachko_2006}. This will be discussed briefly also in Sec.~\ref{sec:towardsdensityfunctionals}.   

\subsection{Extension to the exact embedding projection}
\label{sec:ExactEmbeddingExtension}

With regard to the accuracy of projecting the interacting problem with a non-interacting projector we want to highlight one specific detail. Since we solve for the ground state in the subspace $\mathcal{E}$, we explicitly restrict the CAS in the $M$-particle sector to Slater determinants that all share the same $(M-2n)$ occupied orbitals. These ''frozen'' orbitals form $\ket{\tilde{K}}$. We expect that the thus constructed approximate interacting ground state is not very accurate if $2n$ is small compared to $M$. It is expected that for a more accurate approximation to the interacting ground state one needs to be close to $2n=M$.   \par

Of course, even in the case that $2n=M$ there is no guarantee that the resulting interacting ground-state wave function is well approximated. As discussed above Eq.~\eqref{eq:schmidt}$, 2n\leq N$ (such that the impurity is smaller or equal to the rest of the system). Only upon increasing the dimension of the complete active space to $2n=N$ (which corresponds to impurity being half the system size) one can guarantee to obtain the exact result. For this, however, one needs to adapt the DMET procedure in general and the projection using the CAS as described in Sec.~\ref{sec:embedding_proj_non-int} is not possible anymore. Until now we have assumed that $2n \leq M$ while for $2n=M$ all orbitals contribute to the complete active space and $\ket{\tilde{0}} \equiv \ket{0}$. This implies that without modifications the above procedure only works for $M \geq  N$, where the half-filling case $2n=M=N$ is still captured. Yet for $M<N$ (which is the usual situation in quantum chemistry, since we usually approximate an infinite-dimensional problem $N \rightarrow \infty$ by some finite value for $N$) and $2n>M$, we can no longer use the above introduced procedure, since we can at most define $2M$ orthonormal orbitals by dividing the full lattice into $A$ and $B$. Hence, for $2n>M$ we cannot even resolve the identity on $A$ in this way. In order to allow for an \textit{in principle exact} limit of the DMET procedure with a mean-field projection for $2n>M$ we need to change the construction. The simplest way is to go back to the general form of the projection defined via Eq.~\eqref{eq:projectorFock}. For a single Slater determinant we know from Eq.~\eqref{eq:SimpleNonInteractingSVD} that the rank of the connection matrix is at most $2M$, i.e., only $2M$ of all the $\lambda_{\alpha}$ are non-zero. Hence only a part of the projection onto a $2^{4n}$-dimensional subspace of the Fock space is determined by $\Phi_{ij}$ and the rest is arbitrary. This is why, if we want to control the rest of these dimensions by some self-consistency condition we need to work with multi-determinant mean-field wave functions. And this can only happen if the auxiliary non-interacting system is degenerate. Such a system can of course be engineered, yet becomes rather impractical and again leads to ambiguities. On the one hand, there are many multi-determinant wave functions that lead to the same impurity 1RDM as it is also the case with single determinant wave functions. By choosing the auxiliary non-interacting system to have a degenerate ground-state manifold that contains all of the necessary determinants, these wave functions can be turned into a ground state. Also, each multi-determinant wave function will lead to a different approximate projection. On the other hand, even then the rank of the connection matrix is not necessarily $2n$. So there might be no clear advantage to enforce this self-consistency condition when approaching the exact projection for $2n=N$.

\subsection{Non-interacting $v$-representability: ambiguities in the fixed points}
\label{subsec:wrongprojector}

Let us next consider the influence of the non-interacting $v$-representability problem on the fixed points. To do so, we employ the self-consistency condition of Eq.~\eqref{eq:SelfConsistency} for the special case where we apply the DMET procedure to a non-interacting reference system. While in practice not relevant, since one always solves a non-interacting system numerically exactly, it highlights potential pitfalls that arise due to the non-interacting $v$-representability issue. We will highlight in the following that we can find a fixed point that is an excited state of the target Hamiltonian. Still we see that the self-consistency condition of Eq.~\eqref{eq:SelfConsistency} is fulfilled, i.e. we have an auxiliary Hamiltonian which shares the same impurity 1RDM.

Assume that the target Hamiltonian has the form of Eq.~\eqref{eq:NonInteracting Hamiltonian} and the auxiliary Hamiltonian is given by Eq.~\eqref{eq:AuxiliaryHamiltonian}. But instead of enforcing that $\ket{\Phi}=\hat{\phi}_{M}^{\dagger} \dots \hat{\phi}_{1}^{\dagger} \ket{0}$ and $\ket{\tilde{\Phi}}=\hat{\tilde{\varphi}}_{M}^{\dagger} \dots \hat{\tilde{\varphi}}_{1}^{\dagger} \ket{0}$ share the same impurity 1RDM, we choose that $\ket{\tilde{\Phi}}$ reproduces the impurity 1RDM of $\ket{\Phi'}=\hat{\phi}_{M+1}^{\dagger} \dots \hat{\phi}_{2}^{\dagger} \ket{0}$. That is, it is not the ground state of the Hamiltonian of Eq.~\eqref{eq:NonInteracting Hamiltonian} but an excited state. Furthermore, in the construction that leads to the auxiliary Hamiltonian of Eq.~\eqref{eq:AuxiliaryHamiltonian} we choose all $\varphi_\mu^{B}(z)$ such that 
\begin{align}
    \phi_1(z) \perp \textrm{span}\{ \varphi_1^{A}(z), \dots, \varphi_M^{B}(z) \}.
\end{align}
If $N$ is large enough, i.e., $2N > 2n+M$, this is always possible. The approximate projector $\hat{P}_s$ and its subspace $\mathcal{E}$ then exclude the actual ground state $\ket{\Phi}$ of the $M$-particle sector of the Hamiltonian of Eq.~\eqref{eq:NonInteracting Hamiltonian} and a mininmization leads to $\ket{\Phi'}$ and the corresponding projection $\hat{P}_s'$. This implies that $\hat{P}_s' \hat{H} \hat{P}_s'$ and $\hat{P}_s \hat{\tilde{H}}\hat{P}_s$ share the same impurity 1RDMs and the self-consistency condition of Eq.~\eqref{eq:SelfConsistency} is fulfilled. And instead of $\ket{\Phi}$ we find $\ket{\Phi'}$ at the fixed point (see also Supp.~\ref{app_project} for an explicit example). Realizing that we can easily construct a fixed-point solution that is even further away from $\ket{\Phi}$ by choosing the $\varphi_\mu^{B}(z)$ such that, e.g., all states $\phi_\mu(z)$ of $\ket{\Phi}$ do not appear in $\ket{\Phi'}$ (provided $2N > 2n + 2M$), the self-consistency condition does not automatically imply accuracy. We therefore do not only find multiple fixed points but also the fixed points can be far away from the exact result $\ket{\Phi}$. 

While the example is rather academic, it nicely illustrates a potential pitfall that the non-interacting $v$-representability poses also in the context of the DMET procedure. Here the results of density-functional theories and their mapping theorems can be potentially helpful. We will discuss this point in more detail in Sec.~\ref{sec:towardsuniqueness}. Alternatively, to overcome these ambiguities, the self-consistency condition is adapted or a global iteration is employed instead. We discuss these two options first.

\subsection{Mean-field embedding via embedded one-body reduced density matrix}
\label{subsec:environment1RDM}

The crucial problem of the self-consistency condition of Eq.~\eqref{eq:SelfConsistency} is that it has no unique solution due to the non-interacting $v$-representability problem. There are many non-interacting systems that produce a given impurity 1RDM. So it seems desirable to avoid this ambiguity. One way that is motivated by the numerical instability of the above procedure is to use the (in practice) more stable condition
\begin{align}\label{eq:SelfConsistency2}
    \textrm{min} \; \| \gamma_{\textrm{emb}}^{s} - \gamma_{\textrm{emb}}'\|_2,
\end{align}
where $\gamma_{\textrm{emb}}^{s}(z_1,z_2)$ is the 1RDM of the auxiliary non-interacting system that provides the approximate mean-field projector $\hat{P}_s$, and $\gamma_{\textrm{emb}}'(z_1,z_2)$ is the 1RDM of the projected interacting problem with Hamiltonian $\hat{P}_s \hat{H} \hat{P}_s$ in the respective $M$-particle sector (see also Supp.~\ref{app_project} for an explicit example). If the full projector is used then $z_1$ and $z_2$ are defined on all of $2N$. If instead, as is common practice, we build the projection using the CAS space, some of the bath orbitals (unentangled occupied/cor orbitals) $\varphi_{\mu}(z)$ for $\mu \in \{ 2n+1, \dots, M\}$ are discarded. In this case $z_1$ and $z_2$ correspond to the original lattice sites, only for $z_1$ and $z_2$ in $A$ (see for an example the embedded 1RDM in CAS representation in Eq.~\eqref{app:eq:embedded1} and in spatial representation in Eq.~\eqref{app:eq:embedded2} and then compare with the original Eq.~\eqref{eq:target_1RDM}, which is identical with the one of the full projection).\footnote{This common practice of expressing our Hamiltonian in the CAS subspace corresponds to discarding the chemical-potential term $\Delta \epsilon$ in Eq.~\eqref{eq:FockSubSpaceHamiltonian} and ignoring that $|\tilde 0\rangle$ does correspond to $|\tilde K\rangle$. 
So one effectively uses a $2n$-particle problem to approximate an $M$-particle one.} \par

Which ever way we choose to determine the projection, we first note that we have to slightly modify our DMET procedure, since now also the $\varphi_\mu^{B}(z)$ are determined by the self-consistency condition of Eq.~\eqref{eq:SelfConsistency2}. The exact solution of the minimum condition of Eq.~\eqref{eq:SelfConsistency2} is always zero. However, this leads to the impractical case of a highly degenerate non-interacting system.~\footnote{Given any interacting 1RDM $\gamma_{\textrm{emb}}'(z_1,z_2)$ we can always construct a completely degenerate non-interacting system such that the ground-state solution in the $M$-particle sector is any combination of $M$-particle Slater determinants. This amounts to using multiple degenerate Slater determinants akin to the extension of the DMET procedure for $2n>M$ discussed above. In general this means that we will have to keep all orbitals and thus we might not find any dimensional reduction for the interacting system, which leaves this approach rather impractical.}. Restricting instead to only allow for a single Slater determinant for the case of $2n \leq M$ to construct $\gamma^s_{\textrm{emb}}(z_1,z_2)$  (in which case the minimum of Eq.~\eqref{eq:SelfConsistency2} is non-zero in general~\cite{DET,pDMET,SDE}) will again lead to a large ambiguity. To see this we again consider the case of a non-interacting reference system. If we choose, following the above considerations, an excited state of the reference system and construct an auxiliary Hamiltonian that has a ground-state wave function with a CAS that excludes orbitals appearing in the ground state of the reference system, we have found the minimum ($\gamma^s_{\textrm{emb}}(z_1,z_2)$ = $\gamma'_{\textrm{emb}}(z_1,z_2)$). Yet this is again an undesirable fixed point.

Thus this simple adaptation of the self-consistency condition is not yet enough to avoid potential problems of the non-interacting $v$-representability for 1RDMs.
 
\subsection{Local vs global iterations}

\label{subsec:global}

So far we have considered the situation of one impurity and investigated the ensuing self-consistency. While this is in principle enough, in practice several impurities $A_x$  with $x \in \{1, \dots, I  \}$ that together constitute the full lattice are used. This leads to yet a further large number of possible constructions and iteration procedures with different convergence criteria. It is then usually assumed that iterating locally until convergence and then step successively through all the impurities leads to the same result as when performing the iterations for all the impurities simultaneously~\cite{DMET_molecules1}.

Firstly, even though we can find for every $A_x$ potentially many auxiliary non-interacting Hamiltonian $\tilde{H}^{(1)}_x(z,z')$ that have the same 1RDM (from a non-degenerate ground state) on $A_x$ as the projected interacting problem, there is no procedure that somehow connects all of these auxiliary Hamiltonians and enforces that the interacting and non-interacting projected 1RDMs agree on the full lattice (for a non-degenerate ground state). The reason being, as discussed in Sec.~\ref{subsec:1RDM}, that interacting and non-interacting Hamiltonians cannot share the same 1RDM. Instead, similar to Sec.~\ref{subsec:environment1RDM}, one can try to minimize the difference between the 1RDMs globally. This leads to a completely \textit{degenerate} auxiliary system and in general there is no dimensional reduction. If we further enforce that we only allow for a single Slater determinant we will again find many fixed points. The reasoning is similar to the previous section. We can consider the case of two non-interacting systems on the full lattice, and can construct projectors that single out some excited state of the target system, and then build (following roughly the construction in Sec.~\ref{subsec:wrongprojector}) an auxiliary system that has this state as its ground state (and generates the chosen projection). This underlines that all ambiguities due to the non-interacting $v$-representability that we encountered locally are also present globally.
 
\section{Using density-functional mappings in density-matrix embedding theory: different unique auxiliary systems and projections}
\label{sec:towardsuniqueness}

There are two main reasons for the discussed ambiguities. First, if we allow for a general non-local Hamiltonian of the form of Eq.~\eqref{eq:FockNonInteracting}, different such non-interacting Hamiltonians can have the same ground-state 1RDM. Second, unless we assume total degeneracy (which is rather impractical), a non-interacting system cannot reproduce the full 1RDM of an interacting system. These non-interacting $v$-representability issues are also the reason why there is no Kohn--Sham construction for 1RDM functional theory. A possible way to avoid theses ambiguities is to use the mapping theorems of density-functional theories that indicate that certain observables are representable in an interacting and a non-interacting system uniquely. For instance, instead of working with the 1RDM, we can consider only its diagonal, i.e., the (one-body spin) density. And following the usual mapping theorems we need to do this globally. In this case we can rely on the Hohenberg--Kohn mapping theorems that guarantee that there is only one auxiliary system that generates a specific density. And based on this uniqueness we have a unique auxiliary non-interacting system associated to any interacting one, at least for the global system. While this does not imply that the auxiliary projection is more accurate (for this we would need to consider the norm difference between the exact projection and the auxiliary one), we avoid the above ambiguities and can use this as a unique starting point for refinements.

The trick is thus to restrict to the density $n(z)=\gamma(z,z)$ as well as the form of possible auxiliary systems. So far the auxiliary system allowed for any non-local single-particle Hamiltonian $H^{(1)}(z,z')$, which introduced the above discussed ambiguities. Yet to have the lattice analogue of the Hohenberg--Kohn mapping theorem we need to restrict to
\begin{align}\label{eq:divisionkinetic}
    H^{(1)}(z,z') = T^{(1)}(z,z') + v(z) \delta(z,z'),
\end{align}
where we fix the hopping/kinetic term $T^{(1)}(z,z')$ to the one of the interacting reference system and we only allow to change the (spin-dependent) single-particle potential $v(z)$. We note that the case of finding a projector based on the density together with the restriction to only local potentials is therefore not just a special case of the usual DMET procedure. Firstly, the basic local impurity construction of Sec.~\ref{subsec:arbitraryprojection} is no longer possible. This is because the local potential cannot change the non-local hopping term and hence the density on the impurity depends also on (at least) the bath. So we can only follow the construction presented in Sec.~\ref{subsec:environment1RDM} or directly enforce the same density globally, similar to Sec.~\ref{subsec:global}. Secondly, we avoid the major drawback of having a completely degenerate auxiliary system and do not need to enforce to only allow a single Slater determinant in the minimization. Further, the simple examples for multiple fixed points are ruled out. We (fortunately) lack the flexibility of the non-local auxiliary Hamiltonians.

While the restriction to only the density $n(z)$ has been used and discussed in the DMET literature~\cite{DET}, the ongoing discussion highlights that this case is special. The relation between DMET and density-based embedding theory is similar as the relation between 1RDM functional theory and density-functional theory. They are closely connected, yet call for quite different practical procedures and approximations. The use of auxiliary non-interacting systems in 1RDM functional theory is usually avoided, while in density-functional theory it is very natural and unambiguous. Similarly, the use of a non-interacting auxiliary system for the density-based procedure seems perfectly suited, while a procedure based on the 1RDM can lead to ambiguities as highlighted above. Indeed, borrowing from density-functional theory on a lattice, we know we can uniquely identify an auxiliary non-interacting system $\hat{H}_s[n]$ and its corresponding Kohn--Sham ground state $\Phi[n]$ (with the exact non-interacting projector $P_s[n]$) from which we can (in principle) uniquely construct the exact interacting ground state $\Psi[n]$ and consequently the exact projector $\hat{P}[n]$. And this holds irrespective of the size of the impurity. So, while in the general DMET procedure only increasing the impurity size can improve the reliability and accuracy, in the density-based embedding theory one can make the procedure exact for any impurity size. And similarly to the usual Kohn--Sham approach we can find the exact projection with
\begin{align}
    \hat{P}[n] = \hat{P}_s[n] + \hat{P}_{Hxc}[n],
\end{align}
where $\hat{P}_{Hxc}[n] = \hat{P}[n] - \hat{P}_s[n]$. While this does not have immediate practical consequences, since we do not know how to approximate $\hat{P}_{Hxc}[n]$ and find the standard density-based procedure with setting $\hat{P}_{Hxc}[n] \equiv 0$, it gives an indication how to proceed towards interacting projections. Also, while using the general interacting projector of Eq.~\eqref{eq:projectorFock} leads to the aforementioned practical issues (symmetrization and unknown number of $M$-particle states), the non-interacting projection and its associated subspace $\mathcal{E}$ can be more practically adapted. For instance, one could aim at approximating the correlated $M$-particle states $\ket{A_{\alpha}}\otimes\ket{B_{\alpha}}$ directly from $\mathcal{E}$. In this way one has direct control over symmetry and the number of particles.

Besides the standard density-based functional theories there are also extensions that consider in addition to the density more complex objects, such as the current density or the kinetic-energy density. These objects are all related to parts of the full 1RDM and highlight that besides its diagonal one can potentially influence further parts of the 1RDM in an interacting as well as a non-interacting system. This then leads to new auxiliary non-interacting systems, whose auxiliary projections are potentially a better first guess to the exact projection than just connecting the density. The quantity we look at here specifically is the kinetic-energy density (for a definition of a Hubbard-type of Hamiltonian see Ref.~\cite{keKS} and in a continuum setting Ref.~\cite{SaraPhD} (Chapter 8)). The kinetic-energy density for a Hamiltonian of the type of Eq.~\eqref{eq:ExactInteractingHamiltonian} would be
\begin{align}
K(z_1,z_2)=\langle \Psi |T^{(1)}(z_1,z_2)\hat{c}^{\dagger}_{z_1}\hat{c}_{z_2} \Psi\rangle + c.c.,
\end{align}
where we used the decomposotion of Eq.~\eqref{eq:divisionkinetic}. While this quantity is closely related to the 1RDM, we note that there are two main differences: (i) the  $T^{(1)}(z_1,z_2)$ (which for a Hubbard-type of Hamiltonian amounts to next-neighbour hopping term) is included and (ii) $z_1$ and $z_2$ do not take all the possible values but only the ones that appear in $T^{(1)}(z_1,z_2)$ (for example in the standard Hubbard it will be only the next neighbors that appear). Then one only allows specific non-local potentials (of the same freedom as the interacting ones) by introducing a mean-field Hamiltonian of the type:
\begin{align}
 H^{(1)}_{ke}(z,z') = T^{(1)}_{ke}(z,z') + v_{ke}(z) \delta(z,z').
\end{align}
Thus the target of the DMET procedure could be adapted so that the auxiliary system is constructed in such a way as to reproduce the density $n$ and the kinetic-energy density $K$ of the interacting system. The advantage of the kinetic-energy density with respect to the 1RDM is that it does not suffer from the idempotency issue, i.e. in general a non-interacting system can share the same ground-state kinetic-energy density as an interacting one~\cite{keKS}. However, the second question to make such a procedure well-defined is, whether the mapping between density and kinetic-energy density and local as well as non-local potential is one-to-one, i.e. 
\begin{align}\label{eq:keKSMapping}
    \left(T^{(1)}_{ke}(z,z'), v_{ke}(z)\right)\leftrightarrow \left(K(z,z'), n(z)\right).
\end{align}
The complication in showing that there is such a mapping lies in the fact that we consider a quantity $K(z,z')$ that now includes the external control field $T_{ke}(z,z')$ as well as internal quantities, e.g., in the usual Hubbard case the first off-diagonal of the 1RDM. We therefore no longer have a simple linear structure as in density-functional theory, where external control field $v(z)$ and the internal control objective $n(z)$ are separate entities and are connected via a Legendre-Fenchel transformation~\cite{lieb2002density,penz2019guaranteed}. This also makes the construction of approximations much more complicated. And it is for such problems, where density-functional methods can benefit strongly from the DMET procedure as we will discuss in Sec.~\ref{sec:towardsdensityfunctionals}. Although there is no general answer to the question whether the mapping of Eq.~\eqref{eq:keKSMapping} exists, recent numerical considerations indicate that this might be the case under certain conditions~\cite{keKS}. Hence in analogy to the density-functional based approach, one could apply a kinetic-energy-density based approach where the exact projector is
\begin{align}
    \hat{P}[K,n] = \hat{P}_s[K,n] + \hat{P}_{Hxc}[K,n].
\end{align}
It seems reasonable to assume that, since now the interacting and the non-interacting systems share more properties, also the zeroth order approximation to the mapping, i.e., $\hat{P}_{Hxc}[K,n] \equiv 0$, is more accurate than the one from the density-functional based approach. Following this logic one can try to identify further potential mappings between the interacting and the auxiliary non-interacting system that allow to make both systems more and more alike. For instance, by including a Peierls phase in the hopping, corresponding to an external magnetic field, also the link current becomes potentially controllable~\cite{Karlsson2018}.

Finally, there is yet a different direct way to overcome the ambiguities associated with the 1RDMs. If instead of zero temperature and definite number of particles one considers a (grand-)canonical setting, the inclusion of the entropy in the (grand-)canonical potential allows to reproduce any interacting (grand-)canonical ensemble by a unique non-interacting one~\cite{giesbertz2019one}. The expressions for the non-interacting auxiliary system in the case of the grand-canonical situation are even analytical (see Ref.~\cite{giesbertz2019one} in Sec.~2.4). This is in accordance to recent extensions of DMET to the (grand-)canonical setting~\cite{c_sun}.

\section{Using density-matrix embedding theory in density-functional theories: a novel approximation scheme}
\label{sec:towardsdensityfunctionals}

Up until now we have focused on the DMET procedure and how we can understand certain subtleties connected to the non-interacting $v$-representability from a density-functional perspective. Having realized how the mappings of density-functional approaches appear in DMET provides us with a very interesting possibility. We can use the DMET methodology to directly approximate the interacting-to-non-interacting mapping that is the basis of the Kohn--Sham approach. Instead of indirectly connecting the interacting reference system with the auxiliary non-interacting system via the energy, we instead can directly connect a non-interacting wave function to an approximate interacting wave function that have the same target observable, e.g., in density-functional theory the density.

This idea has been realized in Ref.~\cite{SDE} for the standard case of density-functional theory. From the $v$-representability of the density in both (interacting and auxiliary non-interacting) systems we have 
\begin{align}
    v(z) \stackrel{E\Psi = \hat{H}[v] \Psi}{\longleftrightarrow} n(z) \stackrel{E_s\Phi = \hat{H}_s[v_s] \Phi}{\longleftrightarrow} v_s(z).
\end{align}
We can of course also express this with the help of the exact embedded Hamiltonians $\hat{H}'[v]$ and $\hat{H}_s'[v_s]$, respectively. This mapping directly defines the exact Hxc potential of density-functional theory by
\begin{align}
    v_{Hxc}[n] = v_{s}[n] - v[n].
\end{align}
If we now approximate the embedded Hamiltonians via a self-consistent mean-field projection that makes $n'(z) = n_s(z)$ on the whole lattice we find the approximate mapping 
\begin{align}
    v(z) \stackrel{\Psi'}{\longleftrightarrow} n'(z) \stackrel{\Phi}{\longleftrightarrow} v_s(z),
\end{align}
where the first part $v(z) \rightarrow n'(z)$ and the interacting wave function $\Psi'$ is now only an approximation to $v(z) \rightarrow n(z)$ and $\Psi$. This means that we have now a new interacting mapping $v \rightarrow n'$ that we connect with the non-interacting system and we thus have the approximate Hxc potential
\begin{align}
    v_{Hxc}'[n'] = v_{s}[n'] - v'[n'].
\end{align}
Here we have indicated by $v'$ that we now have a different interacting mapping (since we use the exact mean-field projection the non-interacting mapping $v_s$ is still exact) and with $n'$ that we have in general also a different density at self consistency when compared to the exact density $n$. However, as has been demonstrated in Ref.~\cite{SDE}, by increasing the size of the impurities $A_x$ the difference in density $\|n-n'\| \rightarrow 0$. This implies that we can consistently increase the accuracy of the approximate $v'_{Hxc}$ even for strongly correlated problems. And we have access to an approximate interacting wave functions which allows to approximate many non-trivial observables that are hard to access in normal density-functional theory~\cite{marques2006time,wilken2007momentum}. 

The above described procedure is a novel alternative to the usual way of obtaining density functionals. The common approach is to approximate the energy expression $E[n]-E_s[n]$ and then obtain the corresponding Hxc potential via functional derivative with respect to $n(z)$. However, for the case of certain more complex functional variables like the kinetic-energy density $K(z,z')$, the usual approach via the energy is no longer viable~\cite{keKS}. In this case the above procedure becomes instrumental to go beyond the few simple approximations known. Hence by using the approximate mappings
\begin{align}
    \left(T^{(1)}(z,z'),v(z)\right) \stackrel{\Psi'}{\longleftrightarrow} \left(K'(z,z'),n'(z)\right) \stackrel{\Phi}{\longleftrightarrow} \left(T^{(1)}_{ke}(z,z'),v_{ke}(z)\right),
\end{align}
induced by the approximate projection we find
\begin{align}
    T'^{(1)}_{xc}[K',n'] &= T^{(1)}_{ke}[K',n'] - T'^{(1)}[K'n']\\
    v_{Hxc}'[K',n'] &= v_{ke}[K',n'] - v'[K',n'].
\end{align}
This allows to determine approximately the self-consistent effective hopping term (effective local mass) $T^{(1)} + T^{(1)}_{xc}$ as well as the effective local potential $v + v_{Hxc}$ in the corresponding generalized Kohn--Sham equations 
\begin{align}
    \epsilon_\mu \varphi_\mu(z) = \sum_{z'=1}^{2N} \left[ \; T^{(1)}(z,z') +  T^{(1)}_{xc}([K,n];z,z') + \left(v(z)+v_{Hxc}([K,n];z) \right) \, \delta(z,z')  \right] \varphi_{\mu}(z'),
\end{align}
with $K(z,z') = \sum_{\mu=1}^{M} (T^{(1)}(z,z') + T^{(1)}_{xc}([K,n];z,z') ) \varphi_{\mu}^{*}(z)\varphi_\mu(z') + c.c.$ and $n(z) = \sum_{\mu=1}^{M} \varphi_{\mu}^{*}(z)\varphi_\mu(z)$.

Finally, let us discuss how DMET can be used in density-matrix functional theories to find new approximation schemes. In all the above cases we do not only have access to the reduced variable under investigation, i.e., the density or the kinetic-energy density, but more importantly to an approximate interacting wave function $\Psi'$. With this we also have access to approximate interacting 1RDMs and two-body reduced density matrices (2RDMs). In this regard the DMET procedure provides a direct approximation to parts of the 2RDM and the corresponding interaction energy as well as to parts of the 1RDM and the corresponding kinetic energies. This is not as trivial as it initially sounds. In 1RDM and 2RDM functional theories it is exceedingly hard to guarantee that a trial density matrix, which is used to minimize the energy functional, corresponds to a physical interacting wave function. This crucial point has several layers of complexity attached. Firstly, while there are simple necessary and sufficient conditions known for a 1RDM to be representable by an ensemble of wave functions (ensemble $N$-representability)~\cite{Coleman}, for the 2RDM these conditions are infeasible in practice~\cite{PhysRevLett.98.110503} and one hence uses rather only a subset~\cite{PhysRevLett.108.263002}. To restrict the search space to only pure states, also for the 1RDM the conditions for pure-state $N$-representability become infeasible in practice~\cite{Klyachko_2006}. Finally, if we only want to consider states that are due to the solution of an interacting Schr\"odinger-typ equation ($v$-representability) then no specific conditions are known in general. Yet using the DMET procedure we can use directly the approximate interacting 1RDMs and 2RDMs associated with $\Psi'$ to minimize the energy as a functional of the respective reduced density matrices. Such an approach would suggest to adopt the usual DMET update procedure and not necessarily use a self-consistency condition. Specifically, after having made an initial guess for the auxiliary system and the corresponding auxiliary projection $\hat{P}^{(0)}_s$ we obtain approximate 1RDMs and 2RDMs. For instance in the case of a Hubbard system with next-neighbor interaction and Hubbard on-site interaction already small impurities $A_x$ are enough to have access to all the necessary parts of the 1RDM and the 2RDM to calculate the energy. By, for instance, a downhill-simplex method~\cite{press2007numerical} one then finds a modified 2RDM and the corresponding 1RDM that has lower energy. Since we have just done so by hand we are not guaranteed that it really corresponds to a wave function. If we construct a non-interacting system that shares some of the properties of this modified reduced density matrices we can use the resulting projection $\hat{P}^{(1)}_s$ to find new physical reduced density matrices, which potentially have lower energy than the previous physical ones. In this way we can perform a minimization over $v$-representable 1RDMs and corresponding 2RDMs.

\section{Conclusion and Outlook}

In this work we have highlighted how density-matrix embedding theory (DMET) and different density-functional theories can be used to supplement each other. For the simplest setting of one-dimensional finite lattices we have given a detailed review of the basics of DMET, which allowed us to directly connect this method with different density-functional-type theories. Certain ambiguities that appear in the DMET procedure could be traced back to well-known issues such as the non-interacting $v$-representability issue for one-body reduced density-matrix functional theory. This suggested to overcome these problems by employing appropriate mappings of density-functional theories, which guarantee unique auxiliary systems. On the other hand we could show that DMET can be used to approximate the interacting-to-non-interacting mapping fundamental to the Kohn--Sham construction directly, which provides an approximate interacting wave function from which advanced functional observables can be determined. Furthermore, the DMET procedure suggests itself as a new way to devise approximations in reduced density-matrix functional theories.

While our results are geared towards a specific setting and stay on a rather abstract level, we think that they show the potential in combining both approaches to the many-electron problem. The on-the-fly-construction of approximate interacting wave functions provides a novel paradigm in density(-matrix)-functional approximations. While in density-functional theories it is usually an energy expression that is approximated in terms of the functional variable or Kohn--Sham orbitals, we have seen here that DMET allows to approximate directly the interacting-to-non-interacting mapping. Considering the long and arduous history of devising more accurate density-functional approximations that also work for strongly-correlated systems this approach is promising. The approximate interacting wave functions and their reduced density matrices could also overcome in certain situations the drawback of density-matrix functional theories to enforce numerically expensive representability conditions. The main problem in both cases is of course how to treat more realistic many-electron problems in three spatial dimensions. But with the advances in the DMET procedure together with novel inversion schemes for the non-interacting mapping~\cite{PhysRevA.49.2421, ospadov2017improved, nielsen2018numerical, callow2020density} it seems worthwhile to further explore a combination of DMET and density-functional-type theories.


\bibliography{ref.bib}
\newpage

\pagebreak
\widetext
\begin{center}
\textbf{\large Supplemental material for: Approximations based on density-matrix embedding theory for density-functional theories}
\end{center}
\setcounter{equation}{0}
\setcounter{figure}{0}
\setcounter{table}{0}
\setcounter{section}{0}
\setcounter{page}{1}
\makeatletter
\renewcommand{\theequation}{S\arabic{equation}}
\renewcommand{\thefigure}{S\arabic{figure}}
\renewcommand{\bibnumfmt}[1]{[S#1]}
\renewcommand{\citenumfont}[1]{S#1}
\renewcommand{\thesection}{S\Roman{section}}

\section*{Short Guide to this supplemental material}

In this supplement we want to accompany the general discussion of the main text with simple, yet pedagogical examples. While the main text stays on an abstract level, we find it helpful to follow the discussion to a large part with explicit examples. This allows to focus on the essentials of the different ingredients of the DMET procedure, which for the simple systems presented in this supplement boil down to elementary matrix manipulations of relatively small matrices. Further, it allows to highlight further subtle issues, such as the proper anti-symmetrization of the physical wave function in different basis representations or that one cannot approximate the wave function of the original problem without the discarded core orbitals even on the impurity, by explicit calculations. For further convenience all of the presented results can be re-calculated with a publicly available code that can be found on \href{https://github.com/iris-theof/DMET_SCF_appendix}{GitHub}. 

In the following we will consider spinless fermions, i.e., the dimension of the different objects discussed here and in the main text differ by a factor of 2 in various places. In the main text we have always kept this factor explicit. This allows to directly compare with the abstract (spin-dependent) objects in the main text. In Supp.~\ref{sec:ExampleProjections} we start with a five-site example, give a simple non-interacting Hamiltonian and determine the three-particle ground state in different basis representations. We then present the different ways to perform the projections of the exact wave function (Supp.~\ref{app:projectwavefunctions}) and of the Hamiltonian (Supp.~\ref{app:ExactProjection}) to calculate the embedded system. We finally exemplify why for interacting systems only the projection in Fock space is applicable straightforwardly (Supp.~\ref{app:interactingprojection}).

In Supp.~\ref{app:nonuniqueness} we then consider a six-site example and then highlight first that we can find infinitely many different non-interacting Hamiltonians for a given impurity 1RDM. In Supp.~\ref{app_project} we then demonstrate that we can construct arbitrary fixed points of the DMET procedure if only the 1RDM on the impurity are matched.

\section{Exemplification of the different projections, subspaces and projected Hamiltonians}
\label{sec:ExampleProjections}


\subsection{The basic spaces, Hamiltonians and ground-state representations}
\label{appendix_single_many}


Following the construction discussed in Sec.~\ref{sec:single_part_Fock_space} we first set-up the single-particle Hamiltonian. In our case of spinless particles on a five-site lattice the single-particle space is $\mathcal{H}^{1} \cong h_1 \cong \mathbb{C}^{5}$. The single-particle Hamiltonian includes in our case only next-neighbour hopping terms (with zero boundary conditions) and is expressed in the standard sites basis $\ket{i}$ with $i \in \{1, \dots, 5\}$ as
\begin{align}
\label{eq:non-int_h}
\boldsymbol{H}^{(1)} \cong \boldsymbol{h}^{(1)}=
\left(
\begin{array}{ccccc}
0 & -1 & 0 & 0 & 0 \\
-1 & 0 & -1 & 0 & 0 \\
0 & -1 & 0 & -1 & 0 \\
0 & 0 & -1 & 0 & -1 \\
0 & 0 & 0 & -1 & 0 \\
\end{array}
\right).
\end{align}
Diagonalizing the matrix $H^{(1)}(i,j)$ we find five five-dimensional orthonormal eigenvectors $\phi_{\mu}(i)$, which allows to express $\hat{H}^{(1)}(i,j) = \sum_{\mu=1}^{5}\epsilon_\mu \phi_{\mu}(i)\phi_{\mu}^{*}(j)$. Furthermore, they allow us to build the anti-symmetrized $M$-particle space $\mathcal{H}_M^{F}$ of dimension ${5 \choose M}$ by constructing all possible $M$-particle Slater determinants as well as to setup the (non-interacting) $M$-particle Hamiltonians in this space. To be specific, for the three-particle case a Slater determinant in the non-symmetrized three-particle site basis $|i,j,k) = \ket{i} \otimes \ket{j} \otimes \ket{k}$ becomes
\begin{align}\label{eq:distinguishablebasisPhi}
   \tilde{\Phi}(i,j,k) &= (i,j,k|\mu, \nu, \xi \rangle \nonumber \\
   &=\frac{1}{\sqrt{3!}}
   \;\mathrm{det}\left|
\begin{array}{ccc}
\phi_{\mu}(i) & \phi_{\mu}(j) & \phi_{\mu}(k) \\
\phi_{\nu}(i) & \phi_{\nu}(j) & \phi_{\nu}(k) \\
\phi_{\xi}(i) & \phi_{\xi}(j) & \phi_{\xi}(k)  \\
\end{array}
\right|
\end{align}
This is only non-zero if $\mu \neq \nu \neq \xi$, which means we have ${5 \choose 3} = 10$ such wave functions. Alternatively, we can construct the three-particle Slater determinants in the non-symmetrized basis $|i,j,k)$ of the anti-symmetrized site basis $\ket{i',j',k'}$ as
\begin{align}\label{eq:antisymmetrizedspacebasis}
(i,j,k|i',j',k'\rangle =\frac{1}{\sqrt{3!}}
   \;\mathrm{det}\left|
\begin{array}{ccc}
\delta_{i i'} & \delta_{j i'} & \delta_{k i'} \\
\delta_{i j'} & \delta_{j j'} & \delta_{k j'}  \\
\delta_{i k'} & \delta_{j k'} & \delta_{k k'}   \\
\end{array}
\right|.
\end{align}
We therefore find the above Slater determinant in the anti-symmetrized basis with $i<j<k$ as
\begin{align}\label{eq:antisymmetrizedslater}
   \Phi(i,j,k) &= \langle i,j,k|\mu, \nu, \xi \rangle \nonumber \\
   &=
   \;\mathrm{det}\left|
\begin{array}{ccc}
\phi_{\mu}(i) & \phi_{\mu}(j) & \phi_{\mu}(k) \\
\phi_{\nu}(i) & \phi_{\nu}(j) & \phi_{\nu}(k) \\
\phi_{\xi}(i) & \phi_{\xi}(j) & \phi_{\xi}(k)  \\
\end{array}
\right|.
\end{align}    

Let us next introduce the Fock space of the spinless five-site problem. If we sum over all possible Slater determinants from $M=0$ to $M=5$, the dimension of the resulting Fock space is $2^{5}$. Defining the anti-commuting creation and annihilation operators $\{\hat{c}_i,\hat{c}_{j}^{\dagger} \} = \delta_{ij}$, we can create upon acting on the vacuum state $\ket{0}$ an orthonormal basis of $2^{5}$ states. For instance, we have
\begin{align}\label{eq:threeparticlefockstate}
  \hat{c}^{\dagger}_{k} \hat{c}^{\dagger}_{j} \hat{c}^{\dagger}_{i} \ket{0} = \ket{\emptyset}_0 \oplus \ket{\emptyset}_{2} \oplus \ket{i,j,k}_3 \oplus \ket{\emptyset}_{4} \oplus \ket{\emptyset}_5 \equiv \ket{i,j,k},  
\end{align}
where we indicate the null vector in the respective subspace by $\ket{\emptyset}_{M}$ and we have overloaded the symbol $\ket{i,j,k}$ as referring to the three-particle state of Eq.~\eqref{eq:antisymmetrizedspacebasis} as well as to the three-particle Fock state of Eq.~\eqref{eq:threeparticlefockstate}. Dimensionally these two states are different since they belong to different spaces. While $\ket{i,j,k} \in \mathcal{H}^{F}_3$ is a ${5 \choose 3}$-dimensional vector with a single non-zero entry, $\ket{i,j,k} \in \mathcal{F}$ is a $2^5$-dimensional vector with a single non-zero entry. 

Next we construct the many particle Hamiltonian by summing over all $M$-particle Hamiltonians, e.g., the three-particle Hamiltonian reads $\sum_{\mu=1}^{5}\sum_{\nu>\mu}^{5}\sum_{\xi>\nu}^{5}(\epsilon_{\mu} + \epsilon_{\nu} + \epsilon_{\xi})\ket{\mu,\nu,\xi}\bra{\mu,\nu,\xi}$. Expressing the eigenstates in the anti-symmetrized site basis $\ket{i, \dots , k}$ we find the Fock-space Hamiltonian of Eq.~\eqref{eq:antisymmetrizedspacebasis} in the concise form of
\begin{align}
\label{eq:Ham_int}
\hat{H}&=-\sum_{\langle i,j\rangle}(\hat{c}_{i}^{\dagger}\hat{c}_{j}+\hat{c}_{j}^{\dagger}\hat{c}_{i}),
\end{align}
where $\langle i,j \rangle$ indicates summation only over next neighbours. If we then consider the three-particle subspace, the minimal-energy solution is simply
\begin{align}
\ket{\Phi}&=\prod_{\mu=1}^{3}\hat{\phi}^{\dagger}_{\mu}\ket{0}=\hat{\phi}^{\dagger}_{1}\hat{\phi}^{\dagger}_{2}\hat{\phi}^{\dagger}_{3}\ket{0},
\end{align}
where every orbital creation operator is defined by
\begin{align}
\label{eq:orb_local_basis}
 \hat{\phi}^{\dagger}_{\mu}=\sum_{k=1}^{5}\underbrace{\phi_{\mu}(k)}_{= C_{k \mu}}\hat{c}^{\dagger}_{k} .
\end{align}
More compactly this reads as

\begin{align}
\ket{\Phi}&=\prod_{\mu=1}^{3}\sum_{k=1}^{5}C_{k\mu}\hat{c}^{\dagger}_k\ket{\rm 0},
\label{eq:Phi_init}
\end{align}
where $C_{k\mu}$ are the overlap elements between the two different bases $\braket{k}{\mu} \equiv \phi_{\mu}(k)$. In other words, $C_{k\mu}$ gives the value the orbital $\mu$ has on site $k$, e.g., here we find
\begin{align}
\mathbf{C}=
\left(
\begin{array}{ccc}
\frac{1}{\sqrt{12}} & -0.5 & \frac{1}{\sqrt{3}} \\
0.5 & -0.5 & 0 \\
\frac{1}{\sqrt{3}} & 0    & -\frac{1}{\sqrt{3}} \\
0.5 & 0.5 & 0 \\
\frac{1}{\sqrt{12}} & 0.5  & \frac{1}{\sqrt{3}}
\end{array}
\right).
\label{eq:c_matrix}
\end{align}
The three lowest eigenvalues of \eqref{eq:non-int_h} are ($-\sqrt{3}$, -1,0), so the ground state energy of the three particle system is $E=-\sqrt{3}-1$. We note that $\ket{\Phi}$ as defined in Eq.~\eqref{eq:Phi_init} is defined only in Fock space. Only upon projecting onto the three-particle subspace do we find a wave function of the form of Eq.~\eqref{eq:antisymmetrizedslater}. Yet we in the following denote both wave functions with the same symbol $\ket{\Phi}$ since they correspond to the same physical object just represented in different spaces. Where the difference matters we will comment on it. \par

The issue of different spaces becomes even more clear once we choose to label the above Fock-state basis functions in a specific order similarly to Eq.~\eqref{eq:fock_basis}. For instance we can define
\begin{eqnarray}
|F_1\rangle&=&\ket{0}\nonumber\\
|F_2\rangle&=&\hc^\dagger_{1}\ket{0}\nonumber\\
|F_3\rangle&=& \hc^\dagger_{2}\ket{0}\nonumber\\
..\nonumber\\
|F_{i}\rangle&=&\hc^\dagger_{l}\hc^\dagger_{j}\hc^\dagger_{k}..\ket{0}\nonumber\\
..\nonumber\\
|F_{2^5}\rangle&=&\hc^\dagger_{1} \hc^\dagger_{2} \hc^\dagger_{3} \hc^\dagger_{4} \hc^\dagger_{5}\ket{0}.
\label{eq:fock_basis_example}
\end{eqnarray}
However, the basis functions that will have non-zero contributions to $|\Phi\rangle$ will be only 10, as it is in the three-particle subspace of the whole Fock space. Then the wave function $\ket{\Phi}$ in Fock space can be written as a linear combination of the basis functions $\ket{F_i}$ similarly to Eq.~\eqref{eq:Psi} as
\begin{eqnarray}
 |\Phi\rangle=\sum_{i=1}^{2^5}\Phi_i|F_i\rangle=\sum_{i=1}^{10}\Phi'_i|F'_i\rangle
 \label{eq:phi_sum}
\end{eqnarray}
where we have used the prime to denote only the three-particle basis functions $\ket{F'_i}$ in Fock space:
\begin{eqnarray}
|F'_1\rangle&=&\hc^\dagger_1 \hc^\dagger_2 \hc^\dagger_3 \ket{0}\nonumber\\
|F'_2\rangle&=&\hc^\dagger_{1} \hc^\dagger_2 \hc^\dagger_4 \ket{0}\nonumber\\
|F'_3\rangle&=&\hc^\dagger_{1} \hc^\dagger_2  \hc^\dagger_5 \ket{0}\nonumber\\
|F'_4\rangle&=&\hc^\dagger_{1} \hc^\dagger_3 \hc^\dagger_4 \ket{0} \nonumber\\
|F'_5\rangle&=&\hc^\dagger_{1} \hc^\dagger_3  \hc^\dagger_5 \ket{0}\nonumber\\
|F'_6\rangle&=&\hc^\dagger_{1} \hc^\dagger_{4} \hc^\dagger_5 \ket{0}\nonumber\\
|F'_7\rangle&=& \hc^\dagger_2 \hc^\dagger_{3}\hc^\dagger_4\ket{0}\nonumber\\
|F'_8\rangle&=&\hc^\dagger_{2}\hc^\dagger_{3}  \hc^\dagger_5\ket{0}\nonumber\\
|F'_9\rangle&=&\hc^\dagger_{2}  \hc^\dagger_{4} \hc^\dagger_5\ket{0}\nonumber\\
|F'_{10}\rangle&=&\hc^\dagger_{3} \hc^\dagger_{4} \hc^\dagger_5 \ket{0}
\label{eq:Fock_space_basis}
\end{eqnarray}
In this basis we can find a different expression for the Fock-space Hamiltonian of Eq.~\eqref{eq:Ham_int}, which is implicitly restricted to the three-particle subspace. Diagonalizing the resulting 10x10 matrix with matrix elements $\langle F'_i|\hat{H}| F'_j\rangle$ leads to the following expansion coefficients
$\Phi'_i$ that appear in Eq.~\eqref{eq:phi_sum}:
\begin{align}
\label{eq:Fock_states_coef}
\Phi'_1=\Phi'_{10}=0.10566 \\\nonumber
\Phi'_2=\Phi'_{3}=\Phi'_{6}=\Phi'_{7}=\Phi'_{9}=0.28867\\\nonumber
\Phi'_4=\Phi'_{8}=0.39434\\\nonumber
\Phi'_5=0.5
\end{align}
To see that this agrees with the definition of Eq.~\eqref{eq:Phi_init}, we compare with $C_{k \mu}$. To do so we carry out the sum and the product appearing in Eq.~\eqref{eq:Phi_init} and the wave function is then expressed in the Fock space basis \eqref{eq:Fock_space_basis}. We find that
\begin{align}\label{eq:phi_2}
\Phi'_i=\mathrm{det}\begin{vmatrix}
C_{j,1} & C_{j,2} & C_{j,3}\\
C_{k,1} & C_{k,2} & C_{k,3}\\
C_{l,1} & C_{l,2} & C_{l,3}\\
\end{vmatrix}.
\end{align}
which is the coefficient associated to a basis function $\ket{F'_i}$, with the sites $j,k,l$ occupied. For instance, for $\ket{\Phi'_2}$ we have $j$=1, $k$=2, $l$=4. Carrying out this procedure for all the terms appearing in Eq.~\eqref{eq:Phi_init} one can verify that it is the same wave function as in Eq.~\eqref{eq:phi_sum}. Here we point out that as long as we work with the creation and annihilation operators, which take into account the anti-symmetry by construction, we do not need to anti-symmetrize the coefficients $C_{k \mu}$. Once we have fixed a basis, e.g., $\ket{F_{i}}$, the coefficients need to be anti-symmetrized, e.g., Eq.~\eqref{eq:phi_2}. Moreover, diagonalizing \eqref{eq:Ham_int} in the Fock space basis gives again the same ground-state energy $E$ as the sum of the three lowest orbital energies of \eqref{eq:non-int_h}, i.e. $E=-\sqrt{3}-1$. \par


\subsection{The different projections of the exact wave function}
\label{app:projectwavefunctions}


In the previous part of the supplement we have highlighted the connections between the single-particle, the multi-particle and the Fock-space perspectives. Now we will employ the above different representations of the same physical object, i.e., the three-particle wave function, to perform the exact projection onto an impurity subspace. To do so we choose the impurity to be $A=\{1,2 \}$, i.e., the first two sites. We then want to find a representation of the wave function such that only two orbitals have a contribution on $A$. This representation will then be used to define the exact projected Hamiltonian on a smaller Fock space $\mathcal{E}$.


\subsubsection{Projection via singular-value decomposition in Fock-space basis}
\label{app:svd_fock}


First we show the projection of the exact wave function performed in a Fock-space basis. We will do so via a SVD in the connecting matrix that involves the Fock-space basis functions on the impurity and the corresponding ones on the environment. We can recast our wave function similarly to Eq.~\eqref{eq:split2} in a way that it will comprise of basis functions $\ket{F^A_{i}}$ that belong only to the impurity and $\ket{F^B_{i}}$ that will belong only to the environment. The number of linearly independent $\ket{F^A_{i}}$ that we get is $2^2$, while for $\ket{F^B_{i}}$ is $2^3$. For this we define two new vacua $\ket{0}_A$ and $\ket{0}_B$ and define $\hat{c}_{1}^{\dagger}$ and $\hat{c}_2^{\dagger}$ only on $\ket{0}_A$ (which leads to a Fock space $\mathcal{F}_A$) and accordingly for sites 3, 4 and 5 that constitute $B$ and $\mathcal{F}_B$. By dimensional correspondence we find $\mathcal{F} \cong \mathcal{F}_A \otimes \mathcal{F}_B$. The Fock states of the respective Fock spaces are
\begin{eqnarray}
|F_1^A\rangle&=&\ket{0}_A\nonumber\\
|F^A_2\rangle&=&\hc^\dagger_2\ket{0}_A\nonumber\\
|F^A_3\rangle&=&\hc^\dagger_{1}\ket{0}_A\nonumber\\
|F^A_4\rangle&=&\hc^\dagger_{1}\hc^\dagger_2\ket{0}_A
\label{eq:A_i}
\end{eqnarray}
and
\begin{eqnarray}
|F^B_1\rangle&=& \ket{0}_B\nonumber\\
|F^B_2\rangle&=& \hc^\dagger_3\ket{0}_B\nonumber\\
|F^B_3\rangle&=&  \hc^\dagger_4\ket{0}_B\nonumber\\
|F^B_4\rangle&=& \hc^\dagger_5\ket{0}_B\nonumber\\
|F^B_5\rangle&=& \hc^\dagger_3  \hc^\dagger_4\ket{0}_B\nonumber\\
|F^B_6\rangle&=& \hc^\dagger_3  \hc^\dagger_5\ket{0}_B\nonumber\\
|F^B_7\rangle&=& \hc^\dagger_4  \hc^\dagger_5 \ket{0}_B\nonumber\\
|F^B_8\rangle&=& \hc^\dagger_{3} \hc^\dagger_{4} \hc^\dagger_5\ket{0}_B
\label{eq:B_i}
\end{eqnarray}
Similar to the local creation and annhihilation operators (see Sec.~\ref{sec:single_part_Fock_space}), while the operators in each class $A$ and $B$ anti-commute, operators of different classes commute. So there is no automatic anti-symmetry of combined wave functions, i.e., when representing the three-particle wave function of $\mathcal{F}$ 
\begin{align}\label{eq:split2_example}
\ket{\Phi} = \sum_{i=1}^{2^2}\sum_{j=1}^{2^{3}}\Phi_{i,j} \ket{F^A_i}\otimes \ket{F^B_j}
\end{align}
the coefficients need to take care of the proper symmetry. The induced basis of $\mathcal{F}$ then corresponds to the previously introduced basis of Eq.~\eqref{eq:fock_basis_example}. This means that the coefficients $\Phi_{i,j}$ are identical with the coefficients $\Phi_i$ that appear in Eq.~\eqref{eq:phi_sum}. This means that there will be only 10 non-zero entries of $\Phi_{i,j}$. Recasting now Eq.~\eqref{eq:split2_example} in terms of only its non-zero entries we find that
\begin{align}
\label{eq:wavef_imp_bath}
 \ket{\Phi} =&\overbrace{\Phi_{1,8}}^{\Phi'_{10}}\overbrace{|F^A_1\rangle\otimes \ket{F^B_8}}^{\ket{F'_{10}}}+\overbrace{\Phi_{2,5}}^{\Phi'_{7}}\overbrace{|F^A_2\rangle\otimes \ket{F^B_5}}^{\ket{F'_{7}}}+\nonumber\\
& \overbrace{\Phi_{2,6}}^{\Phi'_{8}}\overbrace{|F^A_2\rangle\otimes \ket{F^B_6}}^{\ket{F'_{8}}}+\overbrace{\Phi_{2,7}}^{\Phi'_{9}}\overbrace{|F^A_2\rangle\otimes \ket{F^B_7}}^{\ket{F'_{9}}}+\nonumber\\
&\overbrace{\Phi_{3,5}}^{\Phi'_{4}}\overbrace{|F^A_3\rangle\otimes \ket{F^B_5}}^{\ket{F'_{4}}}+\overbrace{\Phi_{3,6}}^{\Phi'_{5}}\overbrace{|F^A_3\rangle\otimes \ket{F^B_6}}^{\ket{F'_{5}}}+\nonumber\\
&\overbrace{\Phi_{3,7}}^{\Phi'_{6}}\overbrace{|A_3\rangle\otimes \ket{F^B_7}}^{\ket{F'_{6}}}+\overbrace{\Phi_{4,2}}^{\Phi'_{1}}\overbrace{|F^A_4\rangle\otimes \ket{F^B_2}}^{\ket{F'_{1}}}+\nonumber\\
&\overbrace{\Phi_{4,3}}^{\Phi'_{2}}\overbrace{|A_4\rangle\otimes \ket{F^B_3}}^{\ket{F'_{2}}}+\overbrace{\Phi_{4,4}}^{\Phi'_{3}}\overbrace{|F^A_4\rangle\otimes \ket{F^B_4}}^{\ket{F'_{3}}}
\end{align}
where the matrix elements $\Phi'_i$ are given by Eq.~\eqref{eq:phi_2}.
Having written the wave function of our example in the form of Eq.~\eqref{eq:split2},  we proceed in performing the SVD on the connecting matrix $\mathbf{\Phi}$ with entries $\Phi_{i,j}$ defined just above (all the other entries that this matrix has are zero)
\begin{align}
    \mathbf{\Phi}=\begin{pmatrix}
0 & 0 & 0 & 0 & 0 & 0 & 0 & \Phi'_{10}\\
0 & 0 & 0 & 0 & \Phi'_{7} & \Phi'_{8} & \Phi'_{9} & 0 \\
0 & 0 & 0 & 0 & \Phi'_{4} & \Phi'_{5} & \Phi'_{6} & 0 \\
0 & \Phi'_{1} & \Phi'_{2} & \Phi'_{3} & 0 & 0 & 0 & 0
\end{pmatrix}.
\end{align}
The connecting matrix for our example reads
\begin{align}
    \mathbf{\Phi}=\begin{pmatrix}
0 & 0 & 0 & 0 & 0 & 0 & 0 & 0.10566\\
0 & 0 & 0 & 0 & 0.28868 & 0.39434 & 0.28868 & 0 \\
0 & 0 & 0 & 0 & 0.39434 & 0.5 &  0.28868 & 0 \\
0 & 0.10566 & 0.28868 & 0.28868 & 0 & 0 & 0 & 0
\end{pmatrix}.
\end{align}
Rotating with $\mathbf{U}$ and $\mathbf{V}^{\dagger}$, which are defined in Eq.~\eqref{eq:new_basis_1} and the discussion after it, we obtain the following states on the impurity 
\begin{align}
    \ket{A_1}&=-0.62978\ket{F^A_2}-0.77678\ket{F^A_3} \label{eq:A_1}\\
    \ket{A_2}&=-\ket{F^A_4}\label{eq:A_2}\\
    \ket{A_3}&=-\ket{F^A_1}\label{eq:A_3}\\
    \ket{A_4}&=0.77678\ket{F^A_2}-0.62978\ket{F^A_3} \label{eq:A_4}
\end{align}
and on the bath:
\begin{align}
    \ket{B_1}=&-0.54283\ket{F^B_5}-0.70812\ket{F^B_6}\label{eq:B_1}\\
    &-0.451557\ket{F^B_7} \\\nonumber
    \ket{B_2}=&-0.25056\ket{F^B_2}-0.6846\ket{F^B_3}\\
    &-0.6846\ket{F^B_4}\nonumber\\
    \ket{B_3}=&-\ket{F^B_8}\label{eq:B_3}\\
    \ket{B_4}=&-0.48654\ket{F^B_5}-0.17310\ket{F^B_6}\label{eq:B_4}\\
    &+0.85634\ket{F^B_7}\nonumber\\
    \ket{B_5}=&-0.43646\ket{F^B_2}-0.32517\ket{F^B_3}\\
    &+0.48493\ket{F^B_4}+0.46861\ket{F^B_5}-0.46861\ket{F^B_6}\nonumber\\
    &+0.171523\ket{F^B_7}\nonumber\\
    \ket{B_6}=&-0.64638\ket{F^B_2}-0.08858\ket{F^B_3}\\
    &+0.32517\ket{B_4}-0.46861\ket{F^B_5}+0.46861\ket{F^B_6}\nonumber\\
    &-0.171523\ket{F^B_7}\nonumber\\
    \ket{B_7}=&-0.57351\ket{F^B_2}+0.64638\ket{F^B_3}\\
    &-0.43646\ket{F^B_4}+0.17152\ket{F^B_5}-0.17152\ket{F^B_6}\nonumber\\
    &+0.06278\ket{F^B_7}\nonumber\\
    \ket{B_8}=&\ket{F^B_1}
\end{align}
and a new connecting matrix which is diagonal
\begin{align}
\label{eq:lambda}
\Lambda=
\left(
\begin{array}{cccccccc}
 0.89919 & 0. & 0. & 0. & 0. & 0. & 0. & 0. \\
 0. & 0.4217 & 0. & 0. & 0. & 0. & 0. & 0. \\
 0. & 0. & 0.10566 & 0. & 0. & 0. & 0. & 0. \\
 0. & 0. & 0. & 0.04955 & 0. & 0. & 0. & 0. \\
\end{array}
\right)
\end{align}
The wave function after the SVD reads
\begin{align}
\label{eq:SVD_Phi_Fock}
|\Phi\rangle=    &\sum_{i=1}^4\lambda_i\ket{A_i}\otimes\ket{B_i}\\\nonumber
    =& 0.89919\cdot\overbrace{(-0.62978\ket{F^A_2}-0.77678\ket{F^A_3})}^{\ket{A_1}}\nonumber\\
    &\otimes\overbrace{(-0.54283\ket{F^B_5}-0.70812\ket{F^B_6}-0.45156\ket{F^B_7})}^{\ket{B_1}}\nonumber\\
    &+0.4217\cdot\overbrace{(-\ket{F^A_4})}^{\ket{A_2}}\nonumber\\
    &\otimes\overbrace{(-0.25056\ket{F^B_2}-0.68455\ket{F^B_3}-0.68455\ket{F^B_4})}^{\ket{B_2}}\nonumber\\
    &+0.10566\cdot\overbrace{(-\ket{F^A_1})}^{\ket{A_3}}\otimes\overbrace{(-\ket{F^B_8})}^{\ket{B_3}}\nonumber\\
   &+0.04955\cdot\overbrace{(0.77678\ket{F^A_2}-0.629778\ket{F^A_3})}^{\ket{A_4}}\nonumber\\
    &\otimes\overbrace{(-0.48654\ket{F^B_5}-0.17310\ket{F^B_6}+0.85634\ket{F^B_7})}^{\ket{B_4}}
     \end{align}

If we compare to Eq.~\eqref{eq:wavef_imp_bath}, we see that this is of course the same wave function.


\subsubsection{Projection via singular-value decomposition in the impurity submatrix}
\label{app:svd}


Here we perform the projection via a SVD for the non interacting system on the orbital coefficient matrix of the creation operators. Due to our choice of $A$ we have
\begin{align}
\mathbf{C}^A=\left(
\begin{array}{ccc}
\frac{1}{\sqrt{12}} & -0.5 & \frac{1}{\sqrt{3}} \\
0.5 & -0.5 & 0
\end{array}
\right),
\end{align}
which corresponds to the first two lines of the matrix defined in Eq.~\eqref{eq:c_matrix}.
Performing a SVD on $\mathbf{C}^A$ we get the factorization for its matrix elements
\begin{equation}\label{eq:svd_orb}
{C}^A_{k\mu}=C_{k, \in \mu}(k \in A)=\sum_{k=1}^{2}\sum_{\nu=1}^3 U_{k i}\Lambda_{i\nu} V_{\nu \mu}^{\dagger}
\end{equation}
where $\mathbf{U}$ (size: $2\times2$) and $\mathbf{V}$ (size: $3\times3$) are both orthonormal matrices and $\mathbf{\Lambda}$ is a $2\times3$ matrix with only 2 entries non-zero on the diagonal.
These matrices for our example read:

\begin{align}
\mathbf{U}&=\left(
\begin{array}{cc}
-0.77678 & -0.62978 \\
-0.62978 & 0.77678 \\
\end{array}
\right)\\
\mathbf{\Lambda}&=\left(
\begin{array}{ccc}
0.99317 & 0. & 0. \\
0. & 0.42460 & 0. \\
\end{array}
\right)\\
\mathbf{V}&=\left(
\begin{array}{ccc}
-0.54283 & 0.48654 & -0.68455 \\
0.70812 & -0.17310 & -0.68455 \\
-0.45156 & -0.85634 & -0.25056 \\
\end{array}
\right)
\end{align}
The matrix $\mathbf{V}$ is now the sought-after rotation matrix, which rotates the orbitals into a new basis. In this new basis, only the first two orbitals have overlap with the first two impurity sites.
\begin{align}
\mathbf{\tilde{C}}&=\mathbf{C}\cdot \mathbf{V}\nonumber\\
&=\left(
\begin{array}{ccc}
\frac{1}{\sqrt{12}} & -0.5 & \frac{1}{\sqrt{3}} \\
0.5 & -0.5 & 0 \\
\frac{1}{\sqrt{3}} & 0    & -\frac{1}{\sqrt{3}} \\
0.5 & 0.5 & 0 \\
\frac{1}{\sqrt{12}} & 0.5  & \frac{1}{\sqrt{3}}.
\end{array}
\right)\cdot \left(
\begin{array}{ccc}
-0.54283 & 0.48654 & -0.68455 \\
 0.70812 & -0.17310 & -0.68455 \\
-0.45156 & -0.85634 & -0.25056 \\
\end{array}
\right)\nonumber\\
&=\left(
\begin{array}{ccc}
-0.77147 & -0.26741& 0  \\
-0.62548& 0.32982& 0 \\
-0.05270& 0.77531& -0.25056\\
0.08264&   0.15672& -0.68455\\
-0.06335& -0.44050& -0.68455\\
\end{array}
\right)
\label{eq:c_rot}
\end{align}
As $V$ is a unitary matrix its determinant is one, which results in leaving the Slater determinant of the original wave function unchanged and the new "rotated" orbitals still orthonormal. Thus the impurity-projected representation of Eq.~\eqref{eq:Phi_init} is
\begin{align}\label{eq:phi_env}
\ket{\Phi} &=  \prod_{\mu =1}^{3}\sum_{k=1}^{5} \tilde{C}_{k \mu} \hat{c}^{\dagger}_{k} \ket{0}\\
&=\left(\prod_{\mu=1}^{2}\sum_{k=1}^{5}\tilde{C}_{k\mu}\hat{c}^{\dagger}_k\right)\underbrace{\left(\sum_{k=3}^{5}\tilde{C}_{k3}\hat{c}^{\dagger}_k\right)}_{=\hat{\tilde{\varphi}}^\dagger_3}\ket{0}\nonumber\\
&=\hat{\tilde{\varphi}}^{\dagger}_{1}\hat{\tilde{\varphi}}^{\dagger}_{2}\hat{\tilde{\varphi}}^{\dagger}_{3}\ket{0} \nonumber
\end{align}
We see in this basis that the third orbital is zero on $A$, i.e. the third orbital belongs purely to the environment (or it is an occupied entangled environment orbital as it is called in the DMET literature) and we hence denote it in accordance to Eq.~\eqref{eq:NewCreation} by $\hat{\tilde{\varphi}}^{\dagger}_{3} = \hat{\varphi}^{B,\dagger}_{3}$.

The first two orbitals have still contributions on the impurity and the environment. To construct the missing further two bath orbitals (we should have 3) we remove the part that belongs to the impurity and renormalize the resulting vectors and creation operators as
\begin{align}
\label{eq:def_b}
    \hat{\varphi}^{B,\dagger}_{\mu} &=\sum_{k=3}^5\hat{c}^{\dagger}_k \frac{\tilde{C}_{k\mu}}{\sqrt{\sum_{l=3}^5|\tilde{C}_{l\mu}|^2}}=\sum_{k=3}^5\hat{c}^{\dagger}_k \underbrace{\frac{\tilde{C}_{k\mu}}{\| \tilde{\varphi}_{\mu}\|_B}}_{=\varphi^{B}_\mu(k)}.
    \end{align}
The normalization factors that appear in the denominator of Eq.~\eqref{eq:def_b} read
    \begin{align}
   \| \tilde{\varphi}_{1}\|_B= \sqrt{\sum_{l=3}^5|\tilde{C}_{l1}|^2}=0.11670\\\nonumber
    \| \tilde{\varphi}_{2}\|_B=   \sqrt{\sum_{l=3}^5|\tilde{C}_{l2}|^2}=0.90538
    \end{align}
such that
    \begin{align}
    \hat{\varphi}_1^{B,\dagger} &
    =-0.45156\hat{c}^{\dagger}_3+0.70812\hat{c}^{\dagger}_4-0.54283\hat{c}^{\dagger}_5\\
    \hat{\varphi}_2^{B,\dagger} &
    =0.85634\hat{c}^{\dagger}_3+0.17310\hat{c}^{\dagger}_4-0.48654\hat{c}^{\dagger}_5
\end{align}
In a similar manner (see Eq.~\eqref{eq:NewCreation}) we can define the renormalization factors $\|\tilde{\varphi}_1\|_A=0.99317$, $\|\tilde{\varphi}_2\|_A=0.42460$ and the 2 impurity orbitals as

\begin{align}
 \hat{\varphi}_1^{A,\dagger} 
   & =-0.77678\hat{c}^{\dagger}_1-0.62978\hat{c}^{\dagger}_2\\ \nonumber\\
   \hat{\varphi}_2^{A,\dagger}
    &=-0.62978\hat{c}^{\dagger}_1+0.77677\hat{c}^{\dagger}_2
\end{align}
If we now express $\Phi$ in these new orbitals (that are no longer normalized on the full lattice $A+B$ but only on the respective sub-lattices) we find with the corresponding normalization coefficients
\begin{align}
\ket{\Phi}
&=\overbrace{\left(\|\tilde{\varphi}_1\|_A \hat{\varphi}_1^{A,\dagger} + \|\tilde{\varphi}_1\|_B\hat{\varphi}_1^{B,\dagger} \right)}^{\hat{\tilde{\varphi}}^\dagger_1}\nonumber\\
&\cdot \overbrace{\left(\|\tilde{\varphi}_2\|_A \hat{\varphi}_2^{A,\dagger} + \|\tilde{\varphi}_2\|_B\hat{\varphi}_2^{B,\dagger}\right)}^{\hat{\tilde{\varphi}}^\dagger_2}
 \hat{\varphi}^{B,\dagger}\ket{0}\nonumber\\
&= \|\tilde{\varphi}_1\|_A  \|\tilde{\varphi}_2\|_A  \hat{\varphi}_1^{A,\dagger}  \hat{\varphi}_2^{A,\dagger}  \hat{\varphi}_3^{B,\dagger} \ket{0}\nonumber\\
&+\|\tilde{\varphi}_1\|_A  \|\tilde{\varphi}_2\|_B 
 \hat{\varphi}_1^{A,\dagger}  \hat{\varphi}_2^{B,\dagger}  \hat{\varphi}_3^{B,\dagger} \ket{0}\nonumber\\
&+ \|\tilde{\varphi}_1\|_B  \|\tilde{\varphi}_2\|_A 
 \hat{\varphi}_1^{B,\dagger}  \hat{\varphi}_2^{A,\dagger}  \hat{\varphi}_3^{B,\dagger} \ket{0}\nonumber\\
&+ \|\tilde{\varphi}_1\|_B  \|\tilde{\varphi}_2\|_B 
 \hat{\varphi}_1^{B,\dagger}  \hat{\varphi}_2^{B,\dagger}  \hat{\varphi}_3^{B,\dagger}\ket{0}
\label{eq:SVD_non_inter}
\end{align}
Since we have now four terms in accordance to Eq.~\eqref{eq:SVD_Phi_Fock}, we can individually compare. Firstly we find that
\begin{eqnarray}
\|\tilde{\varphi}_1\|_A  \|\tilde{\varphi}_2\|_A=0.42170=\lambda_2
\end{eqnarray}
and the corresponding vectors can be associated as (note the anti-symmetrization)
\begin{align}
\hat{\varphi}_1^{A,\dagger}  \hat{\varphi}_2^{A,\dagger}  \ket{0} &=
\underbrace{(\varphi_{1}^{A}(1) \varphi_{2}^{A}(2)- \varphi_{2}^{A}(1) \varphi_{1}^{A}(2))}_{= -1} \hat{c}^{\dagger}_1 \hat{c}^{\dagger}_2 \ket{0}\nonumber\\
&\equiv  \ket{A_2},
\end{align}
and 
\begin{align}
    \hat{\varphi}_3^{B,\dagger} \ket{0}&= \left(-0.25056 \hat{c}^\dagger_3 -0.68455 \hat{c}^\dagger_4 -0.68455  \hat{c}^\dagger_5 \right) \ket{0}\nonumber\\
      \equiv &\ket{B_2}.
\end{align}
We note here again that the left-hand sides of the above equivalence relations are vectors that are defined on a Fock space of the full lattice, while the right-hand sides are defined on Fock spaces of sub-lattices. Consequently they are not the same vectors since they are defined on dimensionally different spaces, yet they describe the same physical states. This allows us to associate  
\begin{align}
   \|\tilde{\varphi}_1\|_A  \|\tilde{\varphi}_2\|_A \hat{\varphi}_1^{A,\dagger}  \hat{\varphi}_2^{A,\dagger}  \hat{\varphi}_3^{B,\dagger}\ket{0} \equiv \lambda_2 \ket{A_2} \otimes \ket{B_2}  
\end{align}
while dimensionally (and also physically) $ \|\tilde{\varphi}_1\|_A  \|\tilde{\varphi}_2\|_A \hat{\varphi}_1^{A,\dagger}  \hat{\varphi}_2^{A,\dagger}\ket{0} \otimes  \hat{\varphi}_3^{B,\dagger}\ket{0}$ would not make sense. This highlights again the subtleties that arise when mixing different representations of the same physical state. We can then proceed by associating the other states in a similar manner. Since $\|\tilde{\varphi}_1\|_A  \|\tilde{\varphi}_2\|_B =0.89919=\lambda_1 $,
\begin{align}
    \hat{\varphi}_1^{A,\dagger}  \ket{0} &= \left(-0.77677 \hat{c}^{\dagger}_1-0.62978   \hat{c}^{\dagger}_2 \right) \ket{0}\nonumber\\
      & \equiv \ket{A_1}
\end{align}
and 
\begin{align*}
    \hat{\varphi}_2^{B,\dagger}  \hat{\varphi}_3^{B,\dagger} \ket{0}&= \underbrace{\left( \varphi^{B}_{2}(3)\varphi^B_{3}(4) - \varphi^B_{2}(4)\varphi^B_{3}(3)\right)}_{=-0.54283} \hat{c}^{\dagger}_3 \hat{c}^{\dagger}_4 \ket{0}\nonumber\\
     &+\underbrace{\left( \varphi^{B}_{2}(3)\varphi^B_{3}(5) - \varphi^B_{2}(5)\varphi^B_{3}(3)\right)}_{=-0.70812} \hat{c}^{\dagger}_3  \hat{c}^{\dagger}_5 \ket{0}\nonumber\\
    &+\underbrace{\left( \varphi^{B}_{2}(4)\varphi^B_{3}(5) - \varphi^B_{2}(5)\varphi^B_{3}(4)\right)}_{=-0.45156}\hat{c}^{\dagger}_4  \hat{c}^{\dagger}_5\ket{0}\nonumber\\
    &\equiv \ket{B_1}.
\end{align*}
we have $\|\tilde{\varphi}_1\|_A  \|\tilde{\varphi}_2\|_B 
\hat{\varphi}_1^{A,\dagger}  \hat{\varphi}_2^{B,\dagger}  \hat{\varphi}_3^{B,\dagger} \ket{0} \equiv \lambda_1 \ket{A_1}\otimes \ket{B_1}$. Further, since $\|\tilde{\varphi}_1\|_B  \|\tilde{\varphi}_2\|_A =0.04955=\lambda_4 $, 
\begin{align}
    \hat{\varphi}_2^{A,\dagger} \ket{0} = \left(0.77678 \hat{c}^{\dagger}_2 - 0.62978 \hat{c}^{\dagger}_1\right)\ket{0} \equiv \ket{A_4}
\end{align} 
and 
\begin{align*}
    \hat{\varphi}_1^{B,\dagger}  \hat{\varphi}_3^{B,\dagger} \ket{0}&= \underbrace{\left( \varphi^{B}_{1}(3)\varphi^B_{3}(4) - \varphi^B_{1}(4)\varphi^B_{3}(3)\right)}_{=0.48654} \hat{c}^{\dagger}_3 \hat{c}^{\dagger}_4 \ket{0}\nonumber\\
     &+\underbrace{\left( \varphi^{B}_{1}(3)\varphi^B_{3}(5) - \varphi^B_{1}(5)\varphi^B_{3}(3)\right)}_{=0.17309} \hat{c}^{\dagger}_3  \hat{c}^{\dagger}_5 \ket{0}\nonumber\\
    &+\underbrace{\left( \varphi^{B}_{1}(4)\varphi^B_{3}(5) - \varphi^B_{1}(5)\varphi^B_{3}(4)\right)}_{=-0.85634}\hat{c}^{\dagger}_4  \hat{c}^{\dagger}_5\ket{0}\nonumber\\
    &\equiv -\ket{B_4},
\end{align*} 
we have $\|\tilde{\varphi}_1\|_B  \|\tilde{\varphi}_2\|_A 
\hat{\varphi}_1^{B,\dagger}  \hat{\varphi}_2^{A,\dagger}  \hat{\varphi}_3^{B,\dagger} \ket{0} \equiv \lambda_4 \ket{A_4}\otimes \ket{B_4}$. Finally, since $\|\tilde{\varphi}_1\|_B \|\tilde{\varphi}_2\|_B =0.10566=\lambda_3 $,
\begin{align}
    \ket{0} \equiv -\ket{A_3}
\end{align} 
and 
\begin{align*}
    \hat{\varphi}_1^{B,\dagger}  \hat{\varphi}_2^{B,\dagger} \hat{\varphi}_3^{B,\dagger} \ket{0}&= \underbrace{\textrm{det}\begin{vmatrix}
\varphi^{B}_{1}(3) & \varphi^{B}_{1}(4) & \varphi^{B}_{1}(5)\\
\varphi^{B}_{2}(3) & \varphi^{B}_2(4) & \varphi^{B}_{2}(5)\\
\varphi^{B}_{3}(3) & \varphi^{B}_3(4) & \varphi^{B}_{3}(5)\\
\end{vmatrix}}_{=1}\hat{c}^{\dagger}_3 \hat{c}^{\dagger}_4 \hat{c}^{\dagger}_5 \ket{0}\\
    &\equiv \ket{B_3},
\end{align*} 
we have $\|\tilde{\varphi}_1\|_B  \|\tilde{\varphi}_2\|_B 
\hat{\varphi}_1^{B,\dagger}  \hat{\varphi}_2^{B,\dagger}  \hat{\varphi}_3^{B,\dagger} \ket{0} \equiv \lambda_3 \ket{A_3}\otimes \ket{B_3}$.
Thus we have explicitly verified that performing the SVD in the orbital coefficient matrix with a subsequent division in impurity and environment orbitals is equivalent to performing the SVD in the connecting matrix in the Fock-space representation.


\subsection{The different constructions of the exact embedded system}
\label{app:ExactProjection}


Next we construct the exact embedded system. We do so first by using the Fock-space projection and then we use the single-particle projection. While the first is the only possibility for doing the exact projection for an interacting system (see also Supp.~\ref{app:interactingprojection}), the latter approach is the one that is employed in practice and is only exact for non-interacting systems.


\subsubsection{Exact embedded Hamiltonian via the Fock-space projection}
\label{app:exactFockprojection}

The Fock-space projection according to Eq.~\eqref{eq:projectorFock} reads in our case 
\begin{eqnarray}
\hat{P} &= \sum_{\alpha=1}^{4}\sum_{\beta=1}^{4}\ket{A_{\alpha}}\otimes\ket{B_{\beta}}\bra{A_{\alpha}}\otimes\bra{B_{\beta}},
\end{eqnarray}
with $\ket{A_{\alpha}}$ given by Eqs.~\eqref{eq:A_1}-\eqref{eq:A_4} and $\ket{B_{\beta}}$ given by Eqs.~\eqref{eq:B_1}-\eqref{eq:B_4}.
It is instructive, however, to see how this projection looks like using the impurity plus bath orbitals (and also the environment one). Since we have already associated these states with each other in the previous section, we readily can associate 

\begin{align}
\hat{P}&\equiv
\hat{\varphi}^{A,\dagger}_1 \hat{\varphi}^{B,\dagger}_2\hat{\varphi}^{B,\dagger}_3 \ket{0} \bra{0}\hat{\varphi}^{B}_3 \hat{\varphi}^{B}_2\hat{\varphi}^{A}_1\nonumber\\
&+\hat{\varphi}^{A,\dagger}_1 \hat{\varphi}^{B,\dagger}_3 \ket{0} \bra{0}\hat{\varphi}^{B}_3 \hat{\varphi}^{A}_1\nonumber\\
&+\hat{\varphi}^{A,\dagger}_1 \hat{\varphi}^{B,\dagger}_1 \hat{\varphi}^{B,\dagger}_2 \hat{\varphi}^{B,\dagger}_3 
\ket{0} \bra{0}
\hat{\varphi}^{B}_3  \hat{\varphi}^{B}_2 \hat{\varphi}^{B}_1 \hat{\varphi}^{A}_1\nonumber\\
&+\hat{\varphi}^{A,\dagger}_1 \hat{\varphi}^{B,\dagger}_1 \hat{\varphi}^{B,\dagger}_3
\ket{0} \bra{0}
\hat{\varphi}^{B}_3\hat{\varphi}^{B}_1\hat{\varphi}^{A}_1\nonumber\\
&+\hat{\varphi}^{A,\dagger}_1 \hat{\varphi}^{A,\dagger}_2 \hat{\varphi}^{B,\dagger}_2 \hat{\varphi}^{B,\dagger}_3 \ket{0} \bra{0}\hat{\varphi}^{B}_3 \hat{\varphi}^{B}_2 \hat{\varphi}^{A}_2 \hat{\varphi}^{A}_1\nonumber\\
&+\hat{\varphi}^{A,\dagger}_1 \hat{\varphi}^{A,\dagger}_2 \hat{\varphi}^{B,\dagger}_3 
\ket{0} \bra{0}\hat{\varphi}^{B}_3 \hat{\varphi}^{A}_2 \hat{\varphi}^{A}_1\nonumber\\
&+\hat{\varphi}^{A,\dagger}_1 \hat{\varphi}^{A,\dagger}_2 \hat{\varphi}^{B,\dagger}_1 \hat{\varphi}^{B,\dagger}_2 \hat{\varphi}^{B,\dagger}_3 \ket{0} \bra{0}\hat{\varphi}^{B}_3 \hat{\varphi}^{B}_2 \hat{\varphi}^{B}_1 \hat{\varphi}^{A}_2 \hat{\varphi}^{A}_1\nonumber\\
&+\hat{\varphi}^{A,\dagger}_1  \hat{\varphi}^{A,\dagger}_2 \hat{\varphi}^{B,\dagger}_1  \hat{\varphi}^{B,\dagger}_3  \ket{0} \bra{0}\hat{\varphi}^{B}_3  \hat{\varphi}^{B}_1 \hat{\varphi}^{A}_2 \hat{\varphi}^{A}_1\nonumber\\
&+\hat{\varphi}^{B,\dagger}_3 \ket{0} \bra{0}\hat{\varphi}^{B}_3 \nonumber\\
&+\hat{\varphi}^{B,\dagger}_2 \hat{\varphi}^{B,\dagger}_3 \ket{0} \bra{0}\hat{\varphi}^{B}_3\hat{\varphi}^{B}_2 \nonumber\\
&+\hat{\varphi}^{B,\dagger}_1 \hat{\varphi}^{B,\dagger}_2 \hat{\varphi}^{B,\dagger}_3 \ket{0} \bra{0}\hat{\varphi}^{B}_3\hat{\varphi}^{B}_2 \hat{\varphi}^{B}_1 \nonumber\\
&+\hat{\varphi}^{B,\dagger}_1 \hat{\varphi}^{B,\dagger}_3 \ket{0} \bra{0}\hat{\varphi}^{B}_3 \hat{\varphi}^{B}_1\nonumber\\
&+\hat{\varphi}^{A,\dagger}_2  \hat{\varphi}^{B,\dagger}_3  \ket{0} \bra{0}\hat{\varphi}^{B}_3  \hat{\varphi}^{A}_2\nonumber\\
&+\hat{\varphi}^{A,\dagger}_2 \hat{\varphi}^{B,\dagger}_2 \hat{\varphi}^{B,\dagger}_3 \ket{0} \bra{0}\hat{\varphi}^{B}_3 \hat{\varphi}^{B}_2  \hat{\varphi}^{A}_2\nonumber\\
&+\hat{\varphi}^{A,\dagger}_2 \hat{\varphi}^{B,\dagger}_1\hat{\varphi}^{B,\dagger}_2 \hat{\varphi}^{B,\dagger}_3 \ket{0} \bra{0} \hat{\varphi}^{B}_3 \hat{\varphi}^{B}_2 \hat{\varphi}^{B}_1  \hat{\varphi}^{A}_2 \nonumber\\
&+\hat{\varphi}^{A,\dagger}_2  \hat{\varphi}^{B,\dagger}_1 \hat{\varphi}^{B,\dagger}_3 \ket{0} \bra{0}\hat{\varphi}^{B}_3 \hat{\varphi}^{B}_1  \hat{\varphi}^{A}_2
\label{eq:full_projection}
\end{align}
In this projection appear different terms that project to subspaces with a different number of particles. Since the Hamiltonian is particle-number conserving, contributions like $\braket{\varphi_1^{A} \varphi_2^{B} \varphi^{B}_3}{\hat{H} \varphi_1^{A} \varphi_3^B}$ are zero, yet we still have non-zero contributions within different particle-number subspaces. If we do not restrict at this point to only the three-particle subspace, a minimization of the projected Hamiltonian will lead to a different ground state with different number of particles. In this approach we therefore have to \textit{restrict by hand} to those states such that the correct projector becomes  

\begin{eqnarray}
\label{eq:Proj_Fock_1}
\hat{P}'&= \ket{A_1} \otimes  \ket{B_1} \bra{B_1}\otimes \bra{A_1}\nonumber\\
 &+\ket{A_2} \otimes  \ket{B_2} \bra{B_2}\otimes \bra{A_2}\nonumber\\
  &+\ket{A_3} \otimes  \ket{B_3} \bra{B_3}\otimes \bra{A_3}\nonumber\\
  &+ \ket{A_4} \otimes  \ket{B_4} \bra{B_4}\otimes \bra{A_4}\nonumber\\
   &+ \ket{A_1} \otimes  \ket{B_4} \bra{B_4}\otimes \bra{A_1}\nonumber\\
&+\ket{A_4} \otimes  \ket{B_1} \bra{B_1}\otimes \bra{A_4}
\end{eqnarray}
This yields a $6\times6$ Hamiltonian matrix $\hat{P}' \hat{H} \hat{P}'$ (see the jupyter notebook for the explicit matrix). Diagonalizing provides its lowest eigenvalue as $E=-\sqrt{3}-1$ with the eigenstate $\Phi=-0.89919 \ket{A_1}\otimes \ket{B_1} -0.42170\ket{A_2} \otimes\ket{B_2}  -0.10566 \ket{A_3}\otimes \ket{B_3} -0.04955 \ket{A_4}\otimes\ket{B_4}$. This agrees with Eq.~\eqref{eq:SVD_Phi_Fock}. Alternatively we could have also restricted the projection further to only the states $\ket{A_{\alpha}}\otimes \ket{B_{\alpha}}$ with $\alpha \in \{1,2,3,4\}$ (leading to a $4\times4$ Hamiltonian) and would have found the same ground state.


\subsubsection{Exact embedded Hamiltonian via the single-particle projection}
\label{embedded_system}


Since the above projection needs to be further restricted by hand to only the right particle sector, it is advantageous if one can directly construct the correct projector without further filtering. This can be done for non-interacting systems on the single-particle level as discussed in Sec.~\ref{sec:embedding_proj_non-int}.  Since one uses in practice always a non-interacting projection the following is the standard way in DMET to construct the embedded system. 

We start from the non-interacting Hamiltonian of the full system, i.e., Eq.~\eqref{eq:non-int_h} and construct the non-interacting 1RDM in the site basis from the three lowest energy orbitals (which are the ones that form the Slater determinant $\Phi$). This will be a $5 \times 5$ (the number of sites) matrix
\begin{align*}
    \gamma(i,j)=\langle \Phi|\hat{c}^{\dagger}_j \hat{c}_i|\Phi\rangle=\sum_{\mu=1}^3 C^{T}_{j \mu} C_{i \mu} 
\end{align*}
As this is a non-interacting 1RDM its eigenvalues are 1 or 0. Next we consider a submatrix of this 1RDM which consists only of the sites that belong to the bath $B$. For our example this is the matrix defined above with $i,j \in B \equiv \{3,4,5\}$ 
\begin{align}
    \gamma_{\textrm{env}}(i,j)= \sum_{\mu=1}^3 C^{T}_{j\mu} C_{i\mu} \textrm{  with  } i,j \in B
\end{align}
Diagonalizing this $3\times3$ submatrix gives 
\begin{align*}
    n_1&=0.01362 \equiv \|\tilde{\varphi}_1\|_B^2 \nonumber\\
   \varphi_1^{B}&= -0.45156 \ket{3} +0.70812\ket{4}-0.54283\ket{5}\\
    n_2&=0.81971 \equiv \|\tilde{\varphi}_2\|_B^2 \nonumber\\
    \varphi^{B}_2&= 0.85634\ket{3}+  0.17310\ket{4}-0.48654\ket{5}\\
    n_3&=1 \equiv \|\tilde{\varphi}_3\|_B^2\nonumber\\
   \varphi^{B}_3&= 0.25056\ket{3}+ 0.68455 \ket{4}+0.68455\ket{5}
\end{align*}
where we have introduced the notation $\hat{c}_i^{\dagger}\ket{0}=\ket{i}$.
The orbital $\varphi^{B}_3$ with occupation number $1$ is called in the DMET literature \textit{unentangled occupied environmental orbital}. It agrees with $\tilde{\varphi}_3 \equiv \varphi_{3}^{B}$ from Supp.~\ref{app:svd}. The two orbitals that have eigenvalues (occupations) between $0$ and $1$ are called the \textit{bath orbitals} and agree with the corresponding ones from Sec.~\ref{app:svd}.

Since they have zero contribution on the impurity $A \equiv \{1,2\}$ we need two further states (the size of the impurity) that are non-zero only on the impurity to express a $5 \times 5$ matrix. While we could use $\varphi_1^{A}$ and $\varphi_2^A$ from Sec.~\ref{app:svd}, we can equivalently use
\begin{align*}
    \varphi^{A}_1&= \ket{1},\\
    \varphi^{A}_2&= \ket{2}.
\end{align*}
Discarding the unentangled occupied environmental orbital $\varphi_3^{B}$ that constitutes the vacuum state $\ket{\tilde{0}} = \hat{\varphi}^{B,\dagger}_3 \ket{0}$ of the Fock space $\mathcal{E}$ (see also Sec.~\ref{sec:chemical_potential}), we are left with the 4 orbitals of the complete active space (CAS), i.e., $\varphi_1^{\textrm{CAS}} = \varphi_1^{A}$, $\varphi_2^{\textrm{CAS}} = \varphi_2^{A}$, $\varphi_3^{\textrm{CAS}} = \varphi_1^{B}$ and $\varphi_4^{\textrm{CAS}} = \varphi_2^{B}$. The corresponding $5 \times 4$ CAS matrix $C^{\textrm{CAS}}_{k \mu} \equiv \varphi^{\textrm{CAS}}_{\mu}(k)$ takes the form of Eq.~\eqref{eq:CAS_matrix}. With this the embedded single-particle Hamiltonian becomes
\begin{align}
\label{eq:hs_embed}
\mathbf{h'_s}&=[\mathbf{C}^{\textrm{CAS}}]^T \mathbf{h_s} \mathbf{C}^{\textrm{CAS}} \\ \nonumber
&=\left(
\begin{array}{cccc}
0.0 &-1.0 &0.0& 0.0\\
-1.0& 0.0 &0.45156& -0.856338\\
0.0& 0.45156 & 1.40829 & -0.08973\\
0.0& -0.85634 & -0.08973 & -0.12802\\
\end{array}
\right)
\end{align}
While here we do not gain much in dimensionality, in the case that the impurity is much smaller than the original lattice, this leads indeed to a large reduction. The eigenvalues and eigenvectors of this single-particle embedded Hamiltonian are
\begin{align*}
\epsilon'_1&=-1.37256\\\nonumber
\varphi'_1&=-0.51387\varphi^{\textrm{CAS}}_1
-0.70532\varphi^{\textrm{CAS}}_2\nonumber\\
&+0.09910\varphi^{\textrm{CAS}}_3 -0.47817\varphi^{\textrm{CAS}}_4
\end{align*}
\begin{align*}
\epsilon'_2&=-0.07922\nonumber\\
\varphi'_2&= 0.63451\varphi^{\textrm{CAS}}_1 +0.05027\varphi^{\textrm{CAS}}_2  \nonumber\\
&-0.06164\varphi^{\textrm{CAS}}_3  -0.76881 \varphi^{\textrm{CAS}}_4
\end{align*}
\begin{align*}
\epsilon'_3&=1.0\nonumber\\
\varphi'_3&= -0.5 \varphi^{\textrm{CAS}}_1    +0.5   \varphi^{\textrm{CAS}}_2 \nonumber\\
&-0.62547\varphi^{\textrm{CAS}}_3   -0.32982\varphi^{\textrm{CAS}}_4
\end{align*}
\begin{align*}
\epsilon'_4&=1.73205\nonumber\\
\varphi'_4&=-0.28868\varphi^{\textrm{CAS}}_1+ 0.5 \varphi^{\textrm{CAS}}_2      \nonumber\\
&+0.77147\varphi^{\textrm{CAS}}_3 -0.26741\varphi^{\textrm{CAS}}_4
\end{align*}
Lifting the single-particle Hamiltonian to the Fock-space $\mathcal{E}$ we have (see also Eq.~\eqref{eq:FockSubSpaceHamiltonian})
\begin{align}
	\hat{H}' &= \sum_{\tilde{k}=1}^{4}\sum_{\tilde{k}'=1}^{4} h'_{s}(\tilde{k}, \tilde{k}') \hat{\varphi}^{\mathrm{CAS}, \dagger}_{\tilde{k}} \hat{\varphi}^{\textrm{CAS}}_{\tilde{k}'} + \frac{\Delta \epsilon}{2} \sum_{\tilde{k}=1}^{4} \hat{\varphi}^{\mathrm{CAS}, \dagger}_{\tilde{k}} \hat{\varphi}^{\mathrm{CAS}}_{\tilde{k}}\\
	&= \sum_{\mu=1}^{4} \epsilon'_\mu \hat{\varphi}_{\mu}^{'\dagger} \hat{\varphi}'_{\mu} + \frac{\Delta \epsilon}{2} \sum_{\mu=1}^{4} \hat{\varphi}'^{\dagger}_{\mu} \hat{\varphi}'_{\mu}
\end{align}
Because we have discarded the unentangled occupied environmental orbital the sought-after ground state is given by the lowest two-particle eigenstate of $\hat{H}'$ which leads to $E'=\epsilon_1' + \epsilon_2'=-1.45179$ and $\ket{\Phi'} = \hat{\varphi}_1^{'\dagger} \hat{\varphi}_2^{'\dagger} \ket{\tilde{0}}$. Because in our case $\Delta \epsilon = \braket{\tilde{0}}{\hat{H} \tilde{0}} = E-E'=-1.28026$. For the orbitals we find
\begin{align*}
\varphi'_1&=0.51387 \ket{1}+0.70532\ket{2}+\nonumber\\
&+0.45422\ket{3}+0.01260\ket{4}-0.17885\ket{5},
\end{align*}
\begin{align*}
\varphi'_2&= 0.63451\ket{1}  + 0.05027\ket{2}+ \nonumber\\
&-0.63053\ket{3} 
 -0.17673\ket{4}  
 +0.40752 \ket{5}.
\end{align*}

If we again disregard the unentangled occupied environmental orbital then the resulting Slater determinant is
\begin{align}
    \tilde{\Phi}'(k,l) = \frac{1}{\sqrt{2}}\left(\varphi'_1(k)\varphi'_2(l) - \varphi'_1(l)\varphi'_2(k)  \right).
\end{align}
While it gives the right impurity 1RDM, it does \textit{not} give the wave function even on the impurity. Because the only non-trivial term is $\tilde{\Phi}'(1,2) = -0.298187$ we can compare to, e.g., $\sum_{k=3}^{5}\tilde{\Phi}(1,2,k) = 0.683012$ or some arbitrary combination with $k$ of Eq.~\eqref{eq:distinguishablebasisPhi}. However, if we also use that we know the discarded orbital, i.e., the form of the vacuum state $\ket{\tilde{0}}$, we find instead
\begin{align}
   \tilde{\Phi}(i,j,k) &= (i,j,k|1, 2, 3 \rangle \nonumber \\
   &=\frac{1}{\sqrt{3!}}
   \;\mathrm{det}\left|
\begin{array}{ccc}
\varphi_{1}'(i) & \varphi_{1}'(j) & \varphi_{1}'(k) \\
\varphi_{2}'(i) & \varphi_{2}'(j) & \varphi_{2}'(k) \\
\varphi_{3}'(i) & \varphi_{3}'(j) & \varphi_{3}'(k)  \\
\end{array}
\right|
\end{align}
and have recovered the full wave function.


\subsection{Interacting systems: Why the projection via the impurity submatrix does not work}
\label{app:interactingprojection}


Let us next see explicitly, why for an interacting system only the (less convenient) projection in Fock space works. The simplest wave function that exhibits the main feature of an interacting many-body wave function (multi-reference character) is the linear combination of two Slater determinants. Here we will use two Slater determinants build from orbitals of the non-interacting Hamiltonian of Eq.~\eqref{eq:non-int_h}, 
\begin{align}
\ket{\Phi_1}&=\hat{\phi}^{\dagger}_{1}\hat{\phi}^{\dagger}_{2}\hat{\phi}^{\dagger}_{3}\ket{\rm vac},
\\
\ket{\Phi_2}&=\hat{\phi}^{\dagger}_{1}\hat{\phi}^{\dagger}_{4}\hat{\phi}^{\dagger}_{5}\ket{\rm vac}.
\end{align}
Our model interacting wave function is then
\begin{align}
    \ket{\Psi}=\nu_1 \ket{\Phi_1}+\nu_2\ket{\Phi_2}
    \label{eq:inter_wavef_1}
\end{align}
with (real) $\nu_1^2+\nu_2^2=1$. We can define, similar to the coefficient matrix $\mathbf{C}$ in Eq.~\eqref{eq:c_matrix}, the coefficient matrix of the multi-determinant wave function as
\begin{align}
\mathbf{D}=
\left(
\begin{array}{ccccc}
\frac{1}{\sqrt{12}} & -0.5 & \frac{1}{\sqrt{3}}&-0.5 & \frac{1}{\sqrt{12}}\\
0.5 & -0.5 & 0 &0.5&-0.5\\
\frac{1}{\sqrt{3}} & 0    & -\frac{1}{\sqrt{3}}&0&\frac{1}{\sqrt{3}} \\
0.5 & 0.5 & 0 &-0.5&-0.5\\
\frac{1}{\sqrt{12}} & 0.5  & \frac{1}{\sqrt{3}}&0.5 &  \frac{1}{\sqrt{12}}
\end{array}
\right),
\label{eq:c_matrix_extended}
\end{align}
where the last two columns correspond to $\phi_4$ and $\phi_5$. With this we can determine the 1RDM of the interacting wave function
\begin{align}
    & \gamma(i,j)=\langle \Psi|\hat{c}^{\dagger}_j \hat{c}_i|\Psi\rangle\\
    &=\nu_1^2 \langle \Phi_1|\hat{c}^{\dagger}_j \hat{c}_i|\Phi_1\rangle+
    \nu_2^2 \langle \Phi_2|\hat{c}^{\dagger}_j \hat{c}_i|\Phi_2\rangle \nonumber\\
    &=\nu_1^2 \sum_{\mu=1}^3 D^{T}_{j,\mu} D_{i,\mu} +
    \nu_2^2 ( D^{T}_{j2} D_{i2}+D^{T}_{j4}D_{i4}+D^{T}_{j5}D_{i5}). \nonumber
\end{align}
If we then fix the missing values, e.g., $\nu_1=0.8$ and $\nu_2=0.6$,
we can calculate numerically the 1RDM and determine its environmental submatrix $\gamma^{\textrm{env}}(i,j)$ which would correspond to $i,j \in \{3,4,5 \}$. The eigenvalues and eigenvectors of $\gamma^{\textrm{env}}(i,j)$ are
\begin{align*}
    n_1&= 0.36530\equiv \|\tilde{\varphi}_1\|_B^2 \nonumber\\
   \varphi^{B}_1&= -0.45153\ket{3}+0.67886 \ket{4}-0.57902 \ket{5}
\end{align*}
\begin{align*}
    n_2&=0.59150\equiv \|\tilde{\varphi}_2\|_B^2 \nonumber\\
   \varphi^{B}_2&= 0.88905\ket{3}+
  0.28734\ket{4}
 -0.35641 \ket{5}
\end{align*}
\begin{align*}
       n_3&= 0.81653\equiv \|\tilde{\varphi}_3\|_B^2   \nonumber\\
   \varphi^{B}_3&= 0.07558\ket{3} +
 0.67570 \ket{4}+
 0.73329  \ket{5}
\end{align*}
While before we did go on by discarding the orbital with $n=1$, here we do not find such an unentangled occupied environmental orbital. Thus the procedure that uses the impurity submatrix does in general not work for interacting systems.


\section{Non-interacting $v$-representability issues: non-uniqueness of mean-field projection and of the DMET fixed point}
\label{app:nonuniqueness}

In the following we will demonstrate explicitly in accordance to the non-interacting $v$-representability problem that we do have multiple approximate projections for a given impurity 1RDM (see also general discussion in Sec.~\ref{subsec:arbitraryprojection}). Since this is an integral part of the DMET iteration procedure it is not surprising that we can also show explicitly that we have multiple fixed points as well (see general discussion in Sec.~\ref{subsec:wrongprojector}). As discussed in the main text we need to have enough flexibility in the system to construct the different projections and fixed points. We therefore in this part of the supplement consider a slightly larger grid and take $N=6$. We still consider only $M=3$ spinless fermions.


\subsection{Non-uniqueness of the mean-field projection}

Here we show explicitly non-uniqueness of the approximate projection by constructing two non-interacting system that have the same impurity 1RDM $\gamma^{s}_{\rm imp}(i,j)$ but different orbitals and thus projections. 

We first make a random choice for a non-interacting system. Let us consider a translation invariant Hubbard system, i.e., we have only next-neighbor hopping with periodic boundary conditions:
\begin{align}
    \bm{h}^{s}=\begin{pmatrix}
0& -1    &  0 & 0& 0 &-1\\
-1& 0   &  -1 & 0& 0 &0\\
0& -1   &  0 & -1 & 0 &0\\
0& 0   &  -1 & 0 & -1 &0\\
0& 0   &  0 & -1 & 0 &-1\\
-1& 0   &  0 & 0 & -1 &0\\
\end{pmatrix}
\label{eq:ham_targ_6}
\end{align}

Diagonalizing it leads to six eigenstates $\{\phi_{1}, ..., \phi_{6}\}$, and choosing the three lowest eigenstates leads $\{\phi_{1},\phi_{2},\phi_{3}\}$ with ground-state energy $E=-4$. The 1RDM
\begin{widetext}
 \begin{align}
    \bm{\gamma}^{s}=\begin{pmatrix}
0.5& 0.33333    &  0 & -0.16666& 0 &0.33333\\
0.33333& 0.5   &  0.33333 & 0& -0.16666 &0\\
0& 0.33333   &  0.5 & 0.33333 & 0 &-0.16666\\
-0.16666& 0   &  0.33333 & 0.5 & 0.33333 &0\\
0& -0.166666  &  0 & 0.33333& 0.5 &0.333333\\
0.333333& 0   &  -0.166666 & 0 & 0.333333 &0.5\\
\end{pmatrix}.
\label{eq:target_1RDM}
\end{align}
\end{widetext}
The resulting impurity 1RDM on $A \equiv\{1,2 \}$ is then
\begin{align}
    \bm{\gamma}^{s}_{\rm imp}=\begin{pmatrix}
0.5& 0.33333    \\
0.33333& 0.5    \\
\end{pmatrix}.
\label{eq:gamma_t_imp}
\end{align}
To then construct a different system with the same impurity 1RDM as its three-particle ground state we first diagonalize $\bm{\gamma}^{s}_{\rm imp}$ of Eq.~\eqref{eq:gamma_t_imp} and get the eigenvalues and the eigenvectors of the impurity 1RDM as
\begin{align}
n_1^{\rm imp}&=0.16666 \equiv \| \tilde{\varphi}_1\|_A^2\nonumber\\
\varphi^{A}_1 &=  0.70711 \ket{1}
 -0.70711  \ket{2}.
 \label{eq:imp_occ_1}
\end{align}
\begin{align}
n_2^{\rm imp}&=0.83333 \equiv \|\tilde{\varphi}_2\|_A^2 \nonumber\\
\varphi^{A}_2&= 0.70711 \ket{1}
 +0.70711  \ket{2}
  \label{eq:imp_occ_2}
\end{align}  
Since $B$ is four-dimensional we have four basis functions. We can choose problem adopted ones by just diagonalizing the environment 1RDM of the original $\gamma^{s}(i,j)$ of Eq.~\eqref{eq:target_1RDM} and use two of them to build our CAS space, i.e.,
\begin{align}
    \bm{\gamma}_{\rm env}^{s}=\begin{pmatrix}
  0.5 & 0.33333 & 0 &-0.16666\\
  0.33333 & 0.5 & 0.33333 &0\\
   0 & 0.33333& 0.5 &0.333333\\
 -0.166666 & 0 & 0.333333 &0.5\\
\end{pmatrix}.
\label{eq:env_sub_1RDM}
\end{align}
This leads to
  \begin{align}
n_1&=0\equiv \|\tilde{\varphi}_1\|_B^2\\
\varphi^{B}_1&=
 - 0.31623\ket{3}+0.63246\ket{4} -0.63246\ket{5} +0.31623\ket{6} 
  \label{eq:env_unocc}
\end{align}

  \begin{align}
n_2&=  0.16666 \equiv \|\tilde{\varphi}_2\|_B^2\\   
\varphi^{B}_2&=
  0.63246\ket{3}     
 -0.31623 \ket{4}
 -0.31623 \ket{5}    
  +0.63246 \ket{6} \nonumber
  \label{eq:n_2_env}
  \end{align}
   \begin{align}
  n_3&= 0.83333 \equiv \|\tilde{\varphi}_3\|_B^2 \\
  \varphi^{B}_3&=
 -0.63246\ket{3}    
 -0.31623 \ket{4}+
  0.31623 \ket{5}+     
  0.63246 \ket{6}   
  \label{eq:n_3_env}
  \end{align}
 \begin{align}
 \label{eq:phi_core}
 n_4&= 1.0 \equiv \| \tilde{\varphi}_4\|_B^2\nonumber\\
 \varphi^{B}_4&=
  0.31623\ket{3}     
 + 0.63246  \ket{4}
 + 0.63246     \ket{5}
 + 0.31623 \ket{6}.  
  \end{align}
We can construct the CAS orbitals that would be used in the auxiliary projection of the target Hamiltonian (where for the purpose of the example is not interacting but in a real application one would be interested in interacting Hamiltonians). The first two CAS orbitals can be always chosen as
\begin{align}
    \varphi^{\rm CAS}_1&= \ket{1}, 
  \label{eq:phi_1_cas}
  \\
    \varphi^{\rm CAS}_2&= \ket{2}.
  \label{eq:phi_2_cas}
\end{align}
Alternatively, we could have used the two orbitals $\phi^A$ of the impurity submatrix as we have discussed in the previous part of the supplement. Because the fourth eigenvector of the environment submatrix is discarded in the usual approximate projection (unentangled occupied/core orbital) and the first orbital is perpendicular to the subspace of the three lowest orbitals, we build the other CAS (environmental) orbitals from the remaining orbitals as
\begin{align}
    \varphi^{\rm CAS}_3&=\varphi^{B}_2,
  \label{eq:phi_3_cas}
\\
  \varphi^{\rm CAS}_4&=\varphi^{B}_3.
  \label{eq:phi_4_cas}
\end{align}
While we do not need this CAS orbitals in this section, they will become important in the next. Further, they will show that we get a very different projection when we compare to the CAS from the different Hamiltonian that we construct next.   

\par

If we now take $\varphi_3^{B}$ and $\varphi_4^{B}$ and define
\begin{align}
    \tilde{\varphi}'_1 &= \sqrt{n_1^{\rm imp}} \varphi^{A}_1 + \underbrace{\sqrt{1-n_1^{\rm imp}}}_{=\sqrt{0.83333}} \varphi_1^B
    \\
    \tilde{\varphi}'_2 &= \sqrt{n_2^{\rm imp}} \varphi^{A}_2 + \underbrace{\sqrt{1-n_2^{\rm imp}}}_{=\sqrt{0.16666}} \varphi_4^B
\end{align}
as well as
\begin{align}
    \tilde{\varphi}'_3 &= \varphi^B_2\\
    \tilde{\varphi}'_4 &= \varphi^B_3.
\end{align}
Since we have now four orthogonal vectors we would still need to choose two orthonormal ones to fill up all of the six dimensional space. However, since we only want to construct a Hamiltonian that has the same impurity 1RDM in the ground-state three-particle sector we leave them undefined but instead choose a set of random numbers $\epsilon_1' \leq \epsilon'_2 \leq \epsilon'_3 <  \epsilon'_4 \leq \epsilon'_5 = \epsilon'_6 =0$. For definiteness, we choose $\epsilon_1' = -4$, $\epsilon'_2= -3 $, $\epsilon'_3= -2$, $\epsilon'_4= -1$ and $\epsilon'_5 =\epsilon'_6=0$. With this the new Hamiltonian is 
\begin{widetext}
 \begin{align}
    \bm{h}^{s}_{\rm new}= \sum_{\mu =1}^{4} \epsilon'_{\mu} \tilde{\varphi}'_{\mu}(i)\tilde{\varphi}'^{*}_{\mu (j)} =\begin{pmatrix}
-1.58333 &-0.91667& -0.58333&   0.16667&  -1.16667&  0.08333\\
-0.91667 &-1.58333 & 0.08333 & -1.16667 &  0.16667 &-0.58333\\
-0.58333 & 0.08333 &-1.58333 &  0.76667 & -0.16667 &-0.11667\\
 0.16667 &-1.16667 & 0.76667 & -1.83333 &  1.03333 &-0.16667\\
-1.16667 & 0.16667 &-0.16667 &  1.03333 & -1.83333  &0.76667\\
 0.08333 &-0.58333 &-0.11667 & -0.16667  & 0.76667 &-1.58333\\
\end{pmatrix}
\label{eq:new_Ham}
\end{align}
\end{widetext}
If we diagonalize the Hamiltonian and take the lowest three eigenvectors we find the three-particle ground-state 1RDM
\begin{widetext}
 \begin{align}
    \bm{\gamma}^{s}_{\rm new} =\begin{pmatrix}
0.50000&  0.33333&  0.16667&    0.00000 & 0.33333&
   0.00000\\
 0.33333  & 0.50000  & 0.00000 &  0.33333 & 0.00000&
   0.16667\\
  0.16667 & 0.00000  & 0.50000 & -0.33333 &  0.00000    &
   0.33333\\
   0.00000 &  0.33333 &  -0.33333 & 0.50000 &-0.16667&
   0.00000 \\
  0.33333 & 0.00000 &0.00000 &-0.16667  &0.50000&
  - 0.33333\\
 0.00000 &  0.16667  & 0.33333 & 0.00000 &-0.33333&
   0.50000
\end{pmatrix}
\label{eq:new_1RDM}
\end{align}
\end{widetext}
and the corresponding ground-state energy is $E'=\epsilon'_1+\epsilon'_2+\epsilon'_3 = -9$. By construction the 1RDM agrees on the impurity but the rest is different. Also, the CAS orbitals that are used in the projection will be different. In our case they become (besides the first two that are always the same)
\begin{align}
    \varphi'^{\rm CAS}_3&=\varphi^{B}_4,
  \label{eq:phi_3_cas_prime}
\\
  \varphi'^{\rm CAS}_4&=\varphi^{B}_1.
  \label{eq:phi_4_cas_prime}
\end{align}
in contrast to the ones of Eqs.~\eqref{eq:phi_3_cas} and \eqref{eq:phi_4_cas}. However, $\varphi^{B}_4$ and $\varphi^{B}_1$ were orbitals that did not belong to the original CAS. That means that also the projection constructed from the hamiltonian $\bm{h}^s_{\rm new}$ will look very different from the one of $\bm{h}^{s}$ of Eq.~\eqref{eq:ham_targ_6}. Thus in this example we have highlighted that by the requirement that the impurity 1RDM is the same there is the possibility to construct completely different projections even in a very simple setting where also the target system is non-interacting and we just consider only a few sites.
 
 \subsection{Non-uniqueness of DMET fixed point}
 \label{app_project}

Next we are going to demonstrate that besides the projection also the fixed point is arbitrary and that it can be arbitrarily far away from the ''exact result''. In our case the ''exact result'' is the three-particle ground state of the following ''target'' Hamiltonian
\begin{widetext}
 \begin{align}
     \hat{\bm{h}}^{\rm tar}\equiv \sum_{\mu=1}^6\epsilon^{\rm tar}_{\mu}\phi^{\rm tar}_{\mu}(i)\phi^{\rm tar}_{\mu}(j) \equiv
     \begin{pmatrix}
         0.91667               &-0.583333
       &   0.166667  &
  -0.083333           &
  -0.583333         &   
  -0.833333                \\
  -0.583333  &
   0.916667  &
  -0.833333   &
  -0.583333 &              
  -0.083333 &
   0.166667        \\        
   0.166667           &
  -0.833333           &     
  -0.233333    &                 -0.033333 &       
  -0.633333 &                
   0.566667 \\
  -0.083333      &
  -0.583333                &  
  -0.033333  &         
  -0.683333 &                
   1.016667 &                
  -0.633333                \\
  -0.583333                &
  -0.083333           &
  -0.633333                &
   1.016667                &
  -0.683333               &
  -0.033333           \\
  -0.833333                &
   0.166667                & 
   0.566667                & 
  -0.633333                &      -0.033333       & 
  -0.233333         
\end{pmatrix}
     \label{eq:H_t_new}
 \end{align}
 \end{widetext}
 where we have defined the orthogonal set of eigenfunctions as
 \begin{widetext}
 \begin{align}
 \{ \phi_1^{\rm tar}=\varphi_1^{B},
     \phi_2^{\rm tar}=\phi_1,
     \phi_3^{\rm tar}=\phi_2,
     \phi_4^{\rm tar}=\phi_{3},
     \phi_5^{\rm tar}=\frac{1}{|\phi_4-\langle \phi_1^{\rm tar}|\phi_4 \rangle \phi_1^{\rm tar}|}(\phi_4-\langle \phi_1^{\rm tar}|\phi_4\rangle \phi_1^{\rm tar}),\nonumber\\
      \phi_6^{\rm tar}=\frac{1}{|\phi_5-\langle \phi_1^{\rm tar}|\phi_5 \rangle \phi_1^{\rm tar}|}(\phi_5-\langle \phi_1^{\rm tar}|\phi_5\rangle \phi_1^{\rm tar})+
      \frac{1}{|\phi_5-\langle \phi_5^{\rm tar}|\phi_5 \rangle \phi_5^{\rm tar}|}(\phi_5-\langle \phi_5^{\rm tar}|\phi_5\rangle \phi_5^{\rm tar}
     )  \}    
 \end{align}
 \end{widetext}
and $\{\phi_1,...,\phi_5 \}$ are the five lowest eigenstates of Eq.~\eqref{eq:ham_targ_6}. Further we have chosen
\begin{align*}
    \epsilon_1^{\rm tar}=-2, \epsilon_2^{\rm tar}=-1, \epsilon_3^{\rm tar}=-0.5, \epsilon_4^{\rm tar}=0.5, \epsilon_5^{\rm tar}=1, \epsilon_6^{\rm tar}=2.
\end{align*}
For this Hamiltonian the three-particle ground-state energy is $E^{\rm tar}=-3.5$ and the corresponding 1RDM is
 \begin{widetext}
 \begin{align}
\bm{\gamma}^{\rm tar}=  \begin{pmatrix}
 0.49796 & 0.35483 & 0.02354& -0.16463&-0.02150 & 0.30980\\
 0.35483 & 0.27354 & 0.08537 &-0.02150 & 0.05980 & 0.24796\\
 0.02354 & 0.08537 & 0.32850 & 0.10980 &  0.44796 &  0.00483\\
-0.16463 &-0.02150 & 0.10980  &0.89796& -0.04517 &   0.22354\\
 -0.02150 & 0.05980 & 0.44796 &-0.04517 & 0.67354 &-0.11463\\
 0.30980  &   0.24796 & 0.00483  &0.22354 &-0.11463 & 0.3285 
\end{pmatrix}
\label{eq:gamma_t_new}
\end{align}
\end{widetext}
with the impurity 1RDM as
 \begin{align}
\bm{\gamma}^{\rm tar}_{\rm imp}=  \begin{pmatrix}
 0.49796 &0.35483\\
 0.35483 &0.27354
\end{pmatrix}.
\label{eq:gamma_t_new_imp}
\end{align}

Next we assume that a DMET iteration step led us to an auxiliary Hamiltonian of the form of Eq.~\eqref{eq:ham_targ_6}. So we follow the DMET procedure and determine the CAS of this auxiliary Hamiltonian (see Eqs.~\eqref{eq:phi_1_cas} to \eqref{eq:phi_4_cas}) and define the embedded Hamiltonian
 \begin{align}
\mathbf{h}'^{\rm tar}&=[\mathbf{C}^{\rm CAS}]^T \mathbf{h}^{\rm tar} \mathbf{C}^{\rm CAS}
\label{eq:ham^t_embed_new}\\
&\begin{pmatrix}
 0.00000 &-1.0000 & -0.63246&  0.63246\\
-1.00000 & 0.0000 & -0.63246&  0.63246\\
-0.63246 &-0.63246 & 0.60000&  0.00000\\
 0.63246 &-0.63246 & 0.00000& -0.60000
\end{pmatrix},
\nonumber
\end{align}
with $\mathbf{C}^{\rm CAS}$ the $6 \times 4$ matrix constructed from these orbitals. Diagonalizing this Hamiltonian and keeping only the two lowest (embedded) orbitals we obtain an embedded 1RDM of the target Hamiltonian in the CAS basis as
\begin{align}
\label{app:eq:embedded1}
    \bm{\gamma}'^{\rm tar}_{\rm CAS}&\equiv\sum_{k=1}^2 \varphi_{k}^{\rm emb}(\mu)\varphi_{k}^{\rm emb}(\nu) \\
    &\equiv\begin{pmatrix}
  0.50000&  0.33333 & 0.26352 &0.26352\\
  0.33333&  0.50000 & 0.26352 & 0.26352\\
  0.26352&  0.26352 & 0.16667 &0.00000\\
 -0.26352&  0.26352 & 0.00000      &0.83333\\
  \end{pmatrix} \nonumber
\end{align}
where $\varphi_{\mu}^{\rm emb}$ are the two lowest eigenstates of Eq.~\eqref{eq:ham^t_embed_new}. Transforming the 1RDM into the site basis by
\begin{align}
\label{app:eq:embedded2}
\gamma'^{\rm tar}_{\rm emb}(i,j)=\sum_{\mu,\nu=1}^4 \gamma'^{\rm tar}_{\rm CAS}(\mu,\nu)
\varphi^{\rm CAS}_{\mu}(i)\varphi^{\rm CAS}_{\nu}(j)
\end{align}
leads to
\begin{widetext}
\begin{align}
 \bm{\gamma}'^{\rm tar}_{\rm emb}=\begin{pmatrix}
 0.5    &  0.33333 &0.00000&      -0.16667 & 0.&       0.33333\\
  0.33333 & 0.5     & 0.33333 &0.00000      &-0.16667 &0.00000     \\
 0.00000    &   0.33333&  0.40000 &      0.13333 &-0.20000   &  -0.26667\\
 -0.16667 &0.00000       &0.13333 &  0.1    & -0.06667 &-0.2\\   
  0.      &-0.16667 &-0.2   &  -0.06667  &0.1&      0.13333\\
  0.33333 &0.00000      &-0.26667 &-0.20000      &0.13333  &0.40000    
\end{pmatrix}
\end{align}
\end{widetext}
We notice that the 1RDM $\bm{\gamma}'^{\rm tar}_{emb}$ constructed from the embedded Hamiltonian of Eq.~\eqref{eq:ham^t_embed_new} does not agree with the target 1RDM $\bm{\gamma}^{\rm tar}$ \textit{even} on the impurity $A$. That is, the approximate impurity 1RDM is
\begin{align}
 \bm{\gamma}'^{\rm tar}_{\rm imp}=\begin{pmatrix}
 0.5 &0.33333 \\
 0.33333 &0.5 
\end{pmatrix},
\end{align}
while the ''exact'' impurity 1RDM is given in Eq.~\eqref{eq:gamma_t_new_imp}. Yet it does agree with the 1RDM $\bm{\gamma}_{\rm imp}$ of the auxiliary Hamiltonian of Eq.~\eqref{eq:ham_targ_6} on the impurity. So we have attained the convergence criterion
\begin{align}
    \bm{\gamma}^s_{\rm imp} = \bm{\gamma}'^{\rm tar}_{\rm imp},
\end{align}
and thus our DMET iteration is finished. Besides that we find completely wrong 1RDMs, also the energy estimate is not necessarily good. To demonstrate this we are going to use the following formula to calculate first the energy of the fragment A:
\begin{align}
\label{eq:tot_energy_new}
\epsilon_{f}^{exact}=\sum_{i=1,2,j=1-6}\bm{h}^{tar}_{i,j}\gamma^{tar}(j,i)=-0.55242
\end{align}
where the expression for the fragment energy is taken from \cite{DMET_molecules1} (Eq. (25))
However, because in practice we do not have the correct 1RDM that corresponds to this Hamiltonian available we need to calculate the fragment energy using the embedded 1RDM:
\begin{align}
\label{eq:tot_energy_appr}
\epsilon_{f}^{emb}=\sum_{i=1,2,j=1-6}\bm{h}^{tar}_{i,j}\gamma^{'tar}_{emb}(j,i)=-0.43816
\end{align}
Following the same procedure after adding to embedded 1RDM the environment orbital that we had originally discarded (so as to have three particles)

\begin{align}
    \gamma_{emb}^{tar,tot}(j,i)= \bm{\gamma}'^{\rm tar}_{emb}(j,i)+\phi_4^{*B}(j)\phi^B_4(i)
\end{align}
we obtain the same wrong fragment energy as in \eqref{eq:tot_energy_appr}.

  The reason why we can construct such a ''bad'' fixed point is that we can instead of the ground state of a target Hamiltonian (in our case the three lowest orbitals of the Hamiltonian of Eq.~\eqref{eq:H_t_new}) end up in an excited state (in our case a Slater determinant that excludes the lowest-energy orbitals $\phi_1^{\rm tar}$). We can engineer that by defining an auxiliary system that has the same impurity 1RDM as the excited state and a CAS that excludes the ground state of the system (in our case the CAS of Eqs.~\eqref{eq:phi_1_cas} to \eqref{eq:phi_4_cas} is orthonormal to $\phi_1^{\rm tar} \equiv \varphi_3^B$). We therefore see that by changing the eigenenergies of our auxiliary Hamiltonian in an almost arbitrary fashion as well as by choosing different eigenstates (yet still the CAS needs to be orthonormal to the lowest orbital $\phi_1^{\rm tar} \equiv \varphi_3^B$) we can find even find many auxiliary systems that lead to this ''bad'' fixed point. Moreover, we can of course generate other ''bad'' fixed points by changing the excited state we target and construct the corresponding auxiliary systems.

\end{document}